\documentclass[%
 aip,
 amsmath,amssymb,
 reprint,%
]{revtex4-1}

\usepackage{graphicx}
\usepackage{dcolumn}
\usepackage{bm}

\usepackage[utf8]{inputenc}
\usepackage[T1]{fontenc}
\usepackage{mathptmx}
\usepackage{etoolbox}
\usepackage{epstopdf, epsfig}
\usepackage{amsfonts}
\usepackage{float}
\usepackage{color}
\usepackage{booktabs, makecell, multirow}
\usepackage{subfigure}

\usepackage{placeins}  
\usepackage{hyperref}
\usepackage{multirow} 
\usepackage{lineno}
\usepackage{booktabs,makecell,threeparttable}
\usepackage[capitalise,nameinlink,noabbrev]{cleveref}
\usepackage{tabularray}
\usepackage[normalem]{ulem}
\usepackage{placeins}

\makeatletter
\def\@email#1#2{%
 \endgroup
 \patchcmd{\titleblock@produce}
  {\frontmatter@RRAPformat}
  {\frontmatter@RRAPformat{\produce@RRAP{*#1\href{mailto:#2}{#2}}}\frontmatter@RRAPformat}
  {}{}
}%
\makeatother
\begin{document}



\title{{A Robust Truncated-Domain Approach for Cone--Jet Simulations in Electrospinning and Electrospraying}}


\author{Ghanashyam K. C.}
\affiliation{Department of Mechanical and Aerospace Engineering, Indian Institute of Technology Hyderabad, Kandi 502 284, Sangareddy, Telangana, India}

\author{Satyavrata Samavedi}
\affiliation{Department of Chemical Engineering, Indian Institute of Technology Hyderabad, Kandi 502 284, Sangareddy, Telangana, India}

\author{Harish N Dixit}
\affiliation{Department of Mechanical and Aerospace Engineering, Indian Institute of Technology Hyderabad, Kandi 502 284, Sangareddy, Telangana, India}
\affiliation{Centre for Interdisciplinary Programs, Indian Institute of Technology Hyderabad, Kandi - 502 284, Sangareddy, Telangana, India}

\date{\today}

\begin{abstract}
Direct numerical simulations of electrospinning and electrospraying are computationally demanding due to large-scale separation between the needle and the tip-to-collector distance. The cone-jet mode that occurs in the vicinity of the needle arises from a delicate balance between surface tension, viscous stresses, inertia, and electric stresses. This mode has a central role in determining the subsequent instabilities of the jet and the eventual outcomes on the collector. Truncated-domain simulations offer a viable alternative but depend critically on the accuracy of far-field electrostatic boundary conditions. Existing truncated-domain approaches based on analytical expressions for the electric potential systematically underestimate the electric field near the needle tip and require empirical tuning informed by prior experiments or full-domain simulations, thereby limiting their predictive capability. Here, we present a general truncated-domain framework for electrohydrodynamic (EHD) simulations of the cone-jet mode that avoids these limitations. Our approach exploits inexpensive full-domain electrostatic simulations to obtain the exact electric field and potential distributions near the needle, which are then imposed as boundary conditions in an EHD simulation carried out on a truncated domain. Comparisons with full-domain EHD simulations and experimental data demonstrate that the proposed approach accurately reproduces cone-jet shapes as well as key physical quantities, including electric currents, charge distributions, velocity fields, and Maxwell stresses, while converging at substantially smaller domain sizes. The formulation eliminates tunable parameters, does not require prior knowledge of the cone-jet configuration, and significantly reduces computational cost, providing a reliable and predictive framework for studying electrohydrodynamic cone-jet flows.
\end{abstract}

\maketitle

\section{Introduction}\label{sec:intro}
Electrospinning and electrospraying are widely used techniques for producing sub-micron fibers and particles, respectively, with applications in areas such as drug release, biosensing, and filtration~\cite{ramakrishna2010science,shabafrooz2014electrospun}. Both techniques involve injecting a fluid through a fine metal needle, with typical inner diameters of several hundred microns, while applying a high voltage -- typically several kilovolts -- between the needle and a grounded collector. The strength of the resulting electric field, determined by the applied potential difference and the distance between the needle and collector, is often comparable to or larger than the opposing surface tension forces. When electrostatic forces dominate, the pendant drop at the needle tip deforms into a conical shape known as the \emph{Taylor cone}~\cite{taylor1964disintegration}. The high charge density at the tip of this cone leads to the ejection of a thin jet~\cite{fernandez2007fluid}. As the jet accelerates under the electric field, it undergoes further thinning and may exhibit a variety of instabilities, including Rayleigh (varicose) instability~\cite{zeleny1914electrical} and bending instability~\cite{anton1934process}, before landing on a grounded collector typically placed tens of centimeters away from the needle. The strength of the electric field is controlled by both the applied voltage and the tip-to-collector distance. The overall process can be broadly divided into three distinct zones. The first zone occurs near the needle tip, where a cone forms at low to moderate voltages. The second zone consists of a thin liquid jet emanating from the cone tip, with a transition region connecting the two. In this region, the jet rapidly thins as it accelerates away from the needle. 
In the third zone, the jet may disintegrate into fine droplets due to the Rayleigh instability, undergo bending or whipping instabilities, or split into multiple smaller jets -- a phenomenon known as ``splaying''. The dynamics observed in this region depend on several factors, including fluid rheology, polymer volume fraction, electric field strength, and the presence of free charges in the liquid. Regardless of the instability type in the third zone, the first two zones typically exhibit a stable cone-jet structure, which determines the jet radius and surface charge density prior to the onset of instabilities \cite{ganan2018review,ponce2018steady,dharmansh2016axisymmetric}. These features and the associated instabilities of electrospinning and electrospraying have been discussed extensively in the literature~\cite{reneker2008electrospinning,ji2024electrospinning}; therefore, only a brief overview of the key aspects relevant to the present study is provided here.

Electrospinning and electrospraying can admit several different modes based on the nature of the polymer/solvent system and operating conditions such as voltage and tip-to-collector distance ~\cite{cloupeau1994electrohydrodynamic}. The axisymmetric cone-jet mode - characterized by a stable cone and a single jet – is often preferred in experiments to ensure high-quality fiber or particle production owing to its steady and predictable behavior. Under specific conditions (e.g.,
high voltages), the stable cone-jet mode can transition to a multi-jet mode, in which several thin jets are ejected simultaneously, rendering the process inherently non-axisymmetric. Recent studies~\cite{arunachalam2024establishment,liu2019droplet} have shown that the characteristics of the cone-jet mode directly influence the properties of the resulting fibers or particles. Numerical simulations have therefore become an important tool for studying the cone-jet dynamics, as they provide access to velocity fields, charge distributions, and electric stress that are often inaccessible to experimental measurement. However, most numerical studies are restricted to relatively small tip-to-collector distances or require very large computational domains, making such simulations expensive. In typical experiments, the tip-to-collector distance ($H'$) is of the order of $15$--$20~\mathrm{cm}$, whereas the cone--jet region extends only over $0.6$--$1~\mathrm{cm}$~\cite{arunachalam2024establishment}, resulting in a separation of scales of nearly one order of magnitude. On the other hand, the inner diameter of the needle range from 250 $\mu$m to 600 $\mu$m requiring grid resolutions of $O(1-5)~\mu$m to resolve the flow within the needle. This pronounced disparity in length scales makes full domain simulations prohibitively expensive for realistic experimental configurations.

Hartman \emph{et al.} \cite{hartman1999electrohydrodynamic} developed a numerical model for electrohydrodynamic atomisation, analysing cone-jet formation, jet breakup, and droplet–field interactions, and proposed scaling laws relating the emitted current to flow rate. Higuera \cite{higuera2003flow} investigated the cone-jet transition using numerical and asymptotic methods, showing that when the transition region is small, local current depends only on non-dimensional flow rate and two fluid-dependent parameters. He also identified a recirculation bubble in the transition zone that moves toward the cone with increasing flow rate. Herrada \emph{et al.} \cite{herrada2012numerical} introduced an interface-tracking model assuming interfacial charge only, omitting jet breakup but partially validating their results against experiments and volume-of-fluid (VOF) simulations by L\'{o}pez-Herrera \emph{et al.} \cite{lopez2011charge}. Dastourani \emph{et al.} \cite{dastourani2018physical} employed OpenFOAM to study the influence of flow rate and voltage, solving the coupled electrohydrodynamic equations with a VOF framework. Huh \emph{et al.} \cite{huh2022simulation} proposed improved interpolation schemes for conductivity and permittivity to reduce spurious interfacial charge leakage, enabling more accurate modelling of charge accumulation and cone-jet formation. Mai \emph{et al.} \cite{mai2023numerical} incorporated the effects of liquid wetting and corona discharge in numerical simulations of electrosprays. Lu \emph{et al.}\cite{lu2024numerical} introduced a flux-correction strategy to minimise charge leakage and examined how liquid  properties influence jet pulsation under constant voltage. Yujie Guo \emph{et al.} \cite{guo2024numerical} explored various spray modes - dripping, cone-jet, multi-jet, jetting - by varying voltage, nozzle height, and flow rate, constructing a regime map to delineate transition regimes. Bin He \emph{et al.} \cite{he2025three} developed a three-dimensional model to study varicose and whipping instabilities, demonstrating that conductivity and flow rate control the transition between these modes within a finite voltage range. Collectively, these studies illustrate that electrospinning and electrospraying modelling has advanced significantly, with ongoing efforts aimed at developing robust numerical strategies that reduce computational cost while retaining physical fidelity.

\section{Motivation and objectives}\label{sec:Motivation and objectives}
To effectively control the electrospraying or electrospinning process, it is essential to predict the operating mode under given conditions. Numerical simulations can aid in optimising operating parameters and provide physical insights beyond the reach of experiments. For simulations to serve as a practical tool, they must remain computationally efficient. One strategy adopted in recent studies~\cite{herrada2012numerical} involves employing approximate boundary conditions for the electric field within a truncated computational domain. These boundary conditions are derived from analytical solutions for the electric field distribution around the needle~\cite{jones1971production,ganan1994electrostatic,herrada2012numerical}.
As the spatial extent of the cone–jet is comparable to the needle size, a truncated computational domain is well-suited for simulating the cone–jet region in its entirety.

Herrada \emph{et al.}~\cite{herrada2012numerical} simulated the cone–jet mode within a truncated domain. Since the liquid jet is confined to a narrow region near the axis, the boundary conditions for the fluid (surrounding gas) and pressure may be prescribed using standard far-field conditions. In contrast, electric field lines are not spatially confined and therefore require careful treatment at the truncated boundaries. Herrada \emph{et al.} employed the analytical expression for the electric potential proposed by Jones and Thong~\cite{jones1971production}, originally derived to describe the electric field distribution around a semi-infinite cylinder or line of charges. The accuracy of this analytical expression can be assessed for realistic configurations, such as a finite needle near a collector, by solving the electrostatic problem without approximations. As discussed later in the present work, the Jones–Thong expression underpredicts the electric field near the needle tip, thereby reducing the magnitude of the electrostatic forces. Moreover, the tuning and selection of parameters in the analytical expression require prior knowledge of the cone–jet configuration, which is typically obtained from full-domain simulations or experiments. This reliance on prior information diminishes the predictive capability of Jones–Thong–based truncated-domain approaches and limits their applicability to general configurations or unexplored parameter regimes.

\begin{figure*}[ht]
\centering
\subfigure[]{\includegraphics[trim=0mm 0mm 0mm 0mm, clip, width=0.4\textwidth]{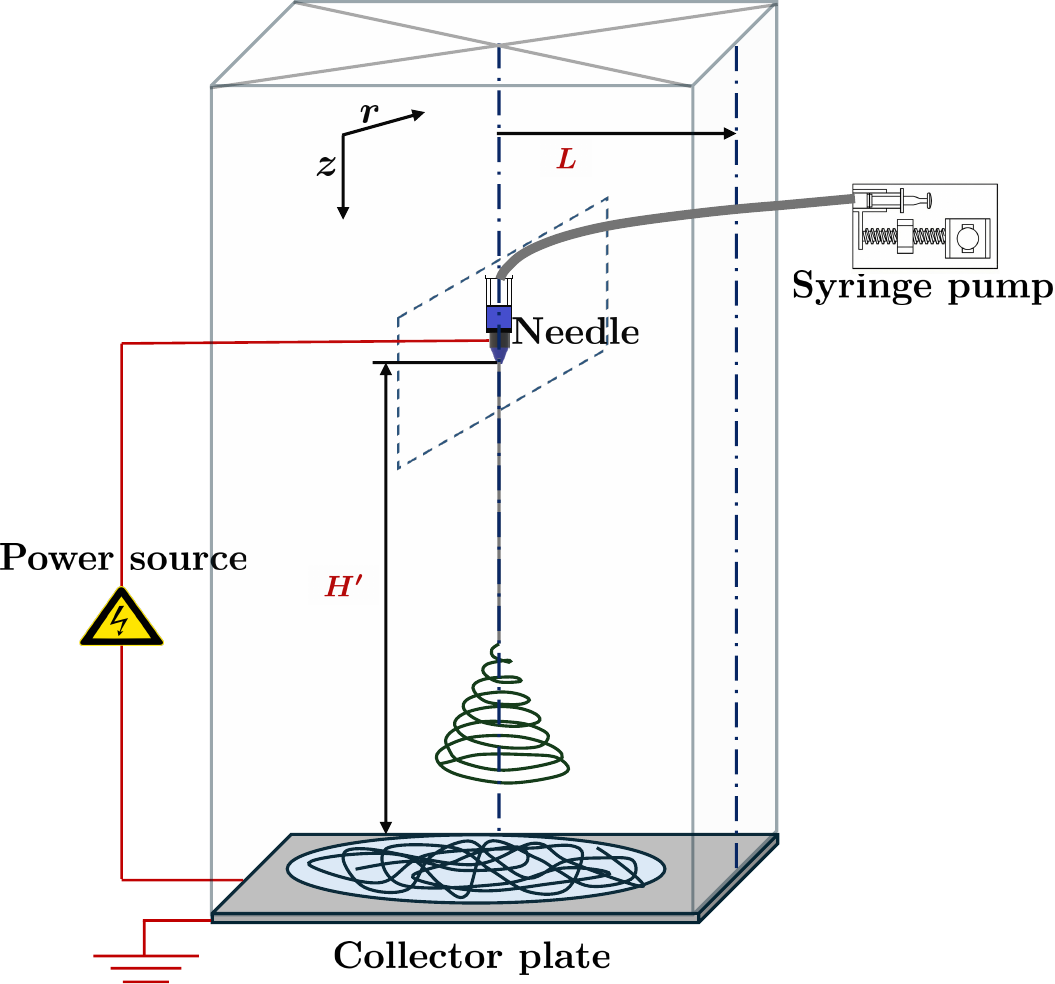}\label{fig:Experimental_setup}}
\subfigure[]{\includegraphics[trim=200mm 50mm 155mm 45mm, clip, width=0.36\textwidth]{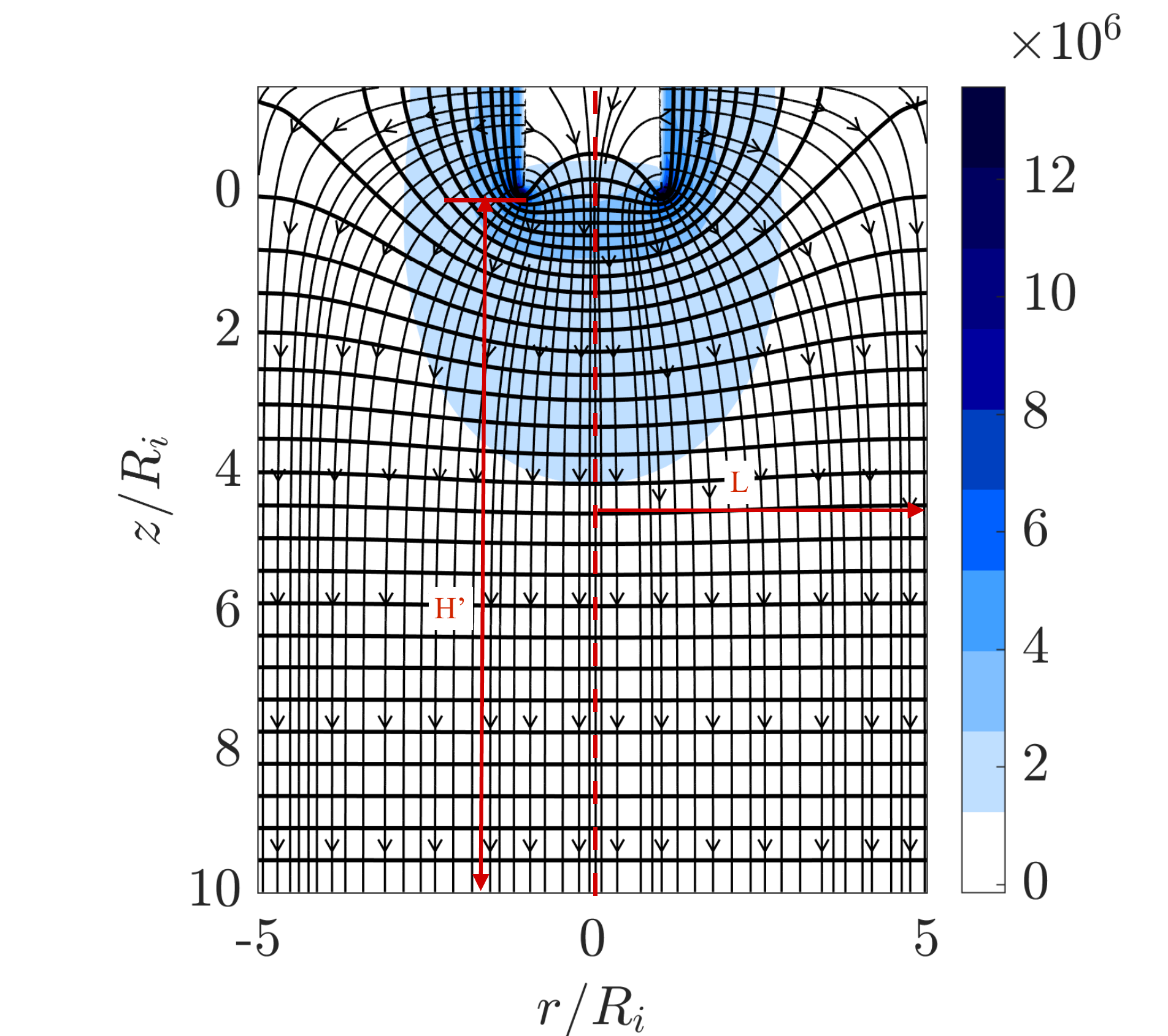}\label{fig:Confinement_effects_1}}
\hspace{5mm}
\subfigure[]{\includegraphics[trim=90mm 50mm 90mm 50mm, clip, width=0.52\textwidth]{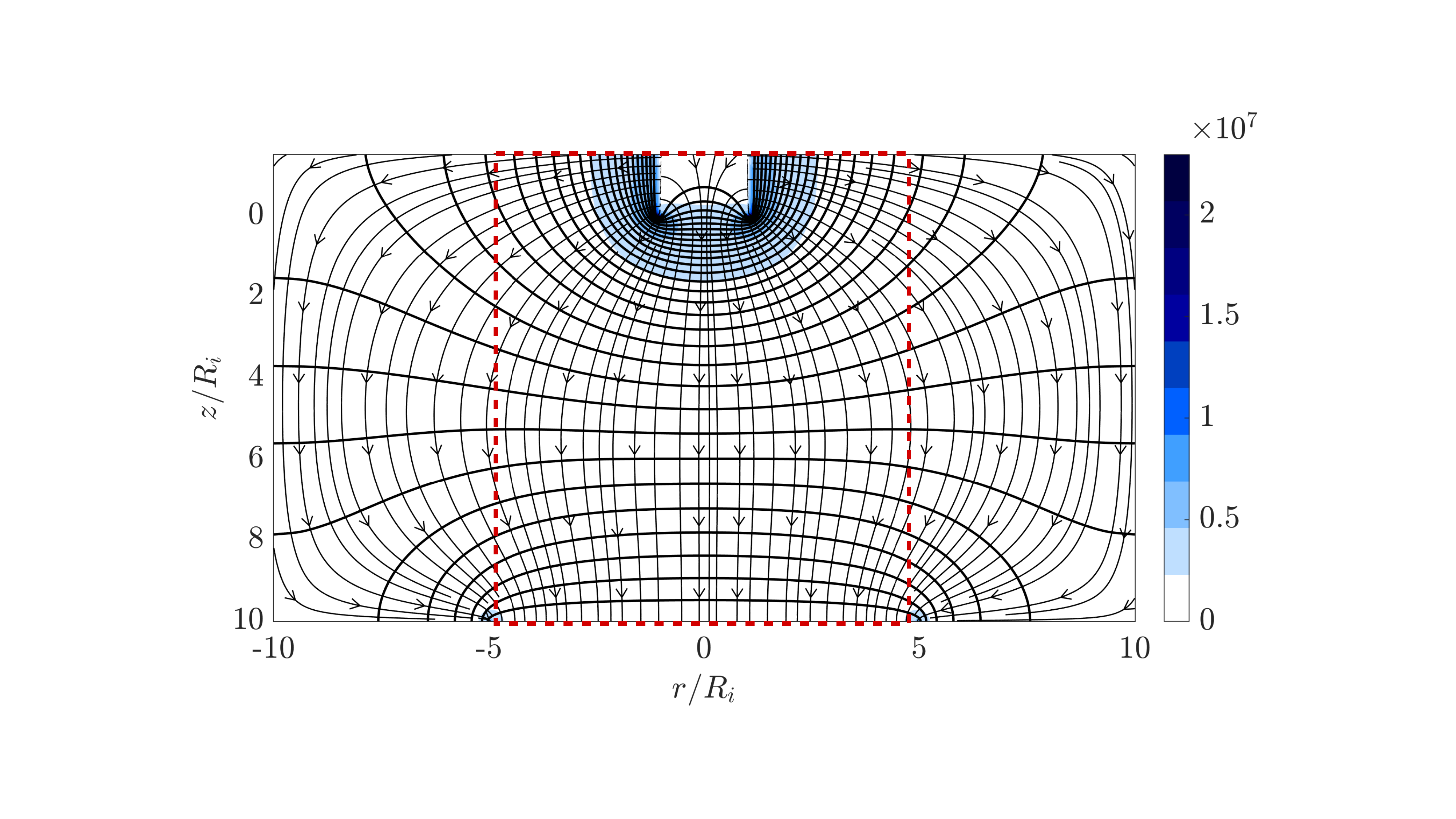}\label{fig:Confinement_effects_2}}
\caption{(a) Schematic of the experimental configuration used in electrospinning and electrospraying, where $L$ denotes the distance from the axis of symmetry of the needle to the lateral confinement wall and $H'$ represents the axial distance from the needle tip to the collector. 
(b) Electric field distribution obtained from purely electrostatic simulations with a lateral boundary distance $L = H'/2$, where $H'$ is the tip-to-collector distance. Electric field lines are shown in black with arrows, while equipotential lines are perpendicular to them.
(c) Electric field distribution for a larger lateral boundary distance $L = H'$, with the collector plate size kept identical to that in panel (b). 
The red box indicates the computational domain size corresponding to $L = H'/2$.
All cases correspond to purely electrostatic simulations with an applied field of $E_\infty = 1.6 \times 10^{6}\,\mathrm{V/m}$.}
\label{fig:Experimental_setup_with_confinement_effects}
\end{figure*}

Despite previous efforts in truncated-domain simulations, several challenges persist in accurately modelling the cone–jet mode in electrospinning and electrospraying. Most experimental configurations employ a blunt-tipped metallic needle through which the liquid flow is regulated using a syringe pump~\cite{arunachalam2024establishment,liu2019droplet}. A high-voltage power supply establishes the required potential difference, with the positive terminal connected to the needle and the grounded terminal attached to a conducting collector plate. A schematic of a typical experimental setup, along with the corresponding electric field distribution, is shown in Fig.~\ref{fig:Experimental_setup_with_confinement_effects}. The apparatus is usually enclosed such that the sidewalls are positioned sufficiently far from the needle, thereby minimising confinement effects. However, because electric fields are highly sensitive to boundary conditions at large distances, care must be taken in numerical simulations to eliminate artificial wall effects. One straightforward approach is to employ a very large computational domain. Nevertheless, resolving the fine interfacial features of the cone–jet requires highly refined meshes, rendering full-domain three-dimensional simulations prohibitively expensive. Figs.~\ref{fig:Confinement_effects_1} and~\ref{fig:Confinement_effects_2} illustrate how lateral confinement alters the electric field distribution near the needle tip, thereby changing the effective field strength for a fixed applied potential and collector geometry. Consequently, understanding the role of sidewalls is essential for obtaining convergent full-domain simulations, an aspect that has received surprisingly little attention in the literature.

The primary objective of this study is to establish a general computational framework for simulating the cone–jet mode in electrospinning and electrospraying within truncated computational domains. The motivation stems from the need to reduce computational cost while retaining accurate predictions of the electric field and interfacial dynamics near the needle, where the cone–jet originates.

The present study proposes an effective strategy to overcome the limitations of Jones and Thong’s analytical formulation by exploiting the relatively low computational cost of purely electrostatic simulations to obtain the correct far-field electric field distribution surrounding the needle. The specific objectives of this work are: (i) to perform full-domain electrostatic simulations to identify the role of lateral boundaries and their influence on the electric field, and (ii) to develop a rational framework for truncated-domain simulations by prescribing boundary conditions derived from these electrostatic solutions.
The remainder of the paper is organised as follows. Section~\ref{sec:problem_setup} describes the problem setup along with the governing equations and boundary conditions. Section~\ref{sec:TDM} presents existing truncated-domain approaches and introduces the proposed method for simulating the cone–jet mode within truncated domains. The main results, together with comparisons to earlier methods, are discussed in Sec.~\ref{sec:results}. Section~\ref{sec:conclusions} summarises the principal findings of the study.

\section{Problem setup}\label{sec:problem_setup}
\begin{figure}[h]
\centering
\includegraphics[trim = 0mm 20mm 20mm 0mm, clip, angle=0, width=0.5\textwidth]{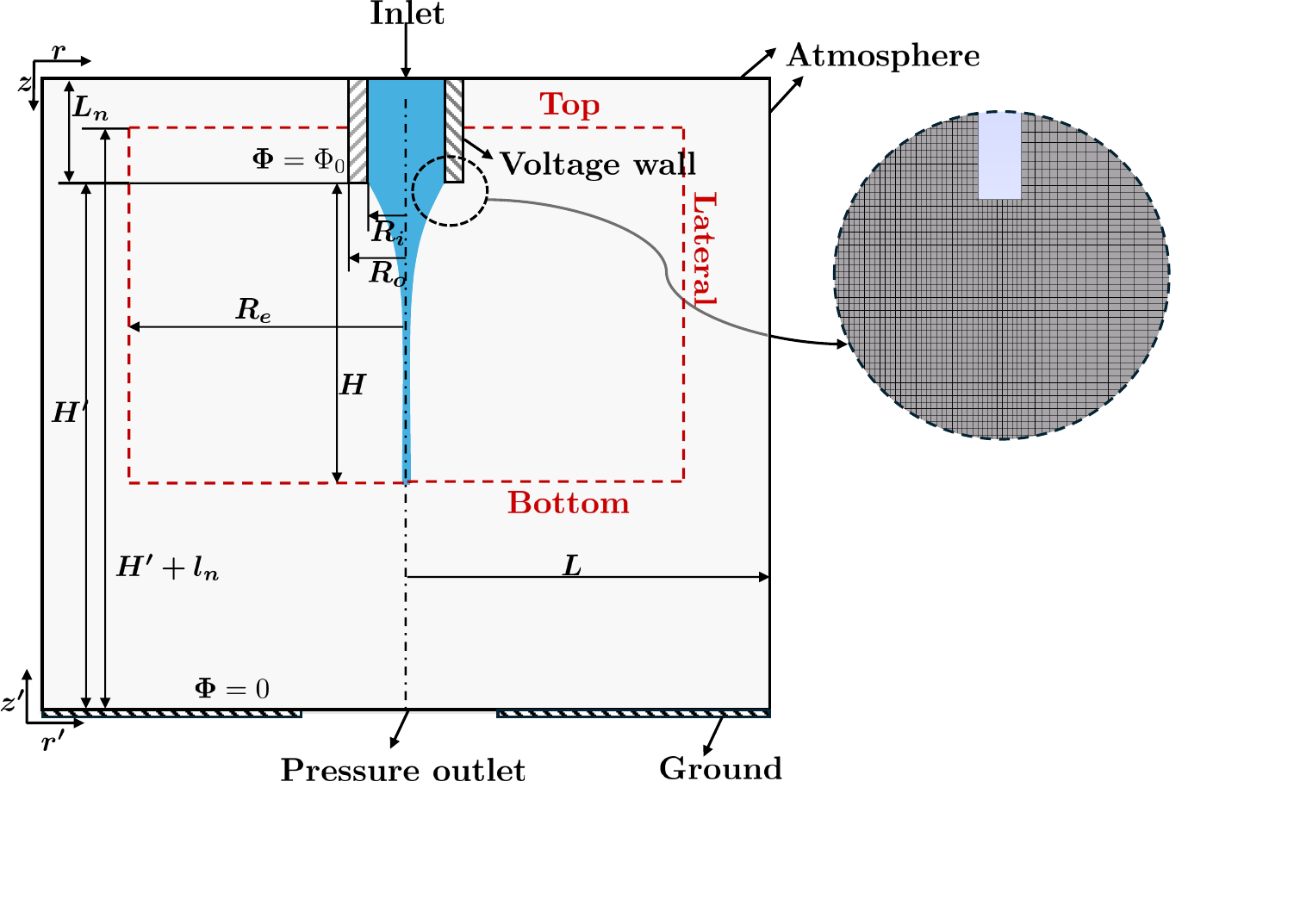}
\caption{Schematic representation of the two computational domains: the full domain (solid black), corresponding to the experimental geometry, and the truncated domain (red dashed). The figure is not to scale. Full-domain simulations (\emph{FDS}) employ standard boundary conditions, whereas truncated-domain simulations (\emph{TDS}) require appropriate boundary conditions on the red dashed boundaries. The local mesh structure is illustrated in the inset.
}
\label{fig:Domain_Schematic}
\end{figure}

A schematic of the computational domain, along with all relevant parameters, is shown in Fig.~\ref{fig:Domain_Schematic}. A Newtonian liquid of viscosity~$\mu$, density~$\rho$, and surface tension~$\gamma$ is injected from a needle of inner radius~$R_i$, outer radius~$R_o$, and length~$L_n$. In addition to gravity~$g$, an axial electric field is imposed by applying a constant potential difference between the needle, held at potential~$\Phi=\Phi_0$, and a grounded collector at~$\Phi=0$. The axial separation between the needle and the collector, referred to as the tip-to-collector distance, is denoted by~$H'$. In most experiments, external boundaries have a negligible influence on the electrospraying or electrospinning process; however, in numerical simulations, the lateral boundary must be placed at a distance~$L$ such that its effect on the solution remains minimal. Insufficient lateral extent can lead to confinement effects that alter the electric field distribution near the needle, as illustrated in Figs.~\ref{fig:Confinement_effects_1} and~\ref{fig:Confinement_effects_2}. As shown later, the appropriate value of~$L$ can be determined systematically through a convergence study. In the present work, the \texttt{interFoam} solver within the OpenFOAM framework is extended to incorporate the Taylor–Melcher leaky-dielectric model~\cite{saville1997electrohydrodynamics}. The resulting solver is capable of simulating two-phase flows with coupled electrostatic effects. The liquid–gas interface is captured using the MULES (Multidimensional Universal Limiter for Explicit Solution) scheme within a volume-of-fluid (VOF) formulation, together with adaptive mesh refinement near the interface to enhance accuracy.

Although the objective of this study is to establish a general framework, the methodology is demonstrated using an axisymmetric configuration for clarity; extension to fully three-dimensional geometries is straightforward. To facilitate comparison with earlier numerical studies~\cite{herrada2012numerical}, 1-octanol is chosen as the working fluid. Its physical properties are: density~$\rho = 827~\mathrm{kg/m^3}$, viscosity~$\mu = 0.0081~\mathrm{kg/(m{\cdot}s)}$, surface tension~$\gamma = 0.0266~\mathrm{N/m}$, electrical conductivity~$\sigma = 9\times10^{-7}~\mathrm{S/m}$, and relative permittivity~$\varepsilon_r = 10$. An adaptive time-stepping procedure is employed to ensure that the Courant number remains below~0.5.
\subsection{Governing equations}
The governing equations are summarized below. The fluid motion is described by the continuity and momentum equations:
\begin{eqnarray}
&\nabla \cdot \mathbf{u} = 0, \label{eq:Continuity}
\end{eqnarray}
\begin{eqnarray}
&\rho\left(\frac{\partial \mathbf{u}}{\partial t} + \mathbf{u}\cdot\nabla\mathbf{u}\right) = -\nabla P + \mu\nabla^2\mathbf{u} + \rho\mathbf{g} + \mathbf{F}_E + \mathbf{F}_{ST}, \label{eq:Navier-stokes}
\end{eqnarray}
where~$\rho$ and~$\mu$ are the density and viscosity, and $\{\mathbf{u},P\}$ represent the velocity and pressure fields. The surface tension force~$\mathbf{F}_{ST}$ is treated as a body force using the CSF model~\cite{brackbill1992continuum}, while~$\mathbf{F}_E$ denotes the electrostatic body force. The surface tension term per unit volume is
\begin{equation}
\mathbf{F}_{ST} = \gamma\kappa\nabla C_{liq} = \gamma(-\nabla\cdot\hat{\boldsymbol{n}})\nabla C_{liq}, \quad
\hat{\boldsymbol{n}} = \frac{\nabla C_{liq}}{|\nabla C_{liq}|+\delta'},
\label{eq:ST_Force}
\end{equation}
where~$\gamma$ is the surface tension,~$\kappa$ the interfacial curvature,~$C_{liq}$ the volume fraction of liquid, and~$\hat{\boldsymbol{n}}$ the interface normal. Density and viscosity are evaluated by weighted arithmetic averaging:
\begin{align}
\rho &= \rho_1 C_{liq} + \rho_2(1 - C_{liq}), \\
\mu &= \mu_1 C_{liq} + \mu_2(1 - C_{liq}).
\label{eq:Interpolation_rho_mu}
\end{align}
The liquid fraction~$C_{liq}$ is obtained from the VoF transport equation:
\begin{equation}
\frac{\partial C_{liq}}{\partial t} + \nabla\cdot(C_{liq}\boldsymbol{u}) + \nabla\cdot[C_{liq}(1 - C_{liq})\boldsymbol{u}_r] = 0,
\label{eq:VOF_eqn}
\end{equation}
where~$\boldsymbol{u}_r$ is an artificial compression velocity used in the MULES scheme to maintain a sharp interface.

The electrostatic body force, obtained from Maxwell’s stress tensor, includes both Coulombic and polarization contributions:
\begin{equation}
\boldsymbol{F}_E = \rho_e\boldsymbol{E} - \frac{1}{2}\nabla(\varepsilon E^2),
\label{eq:Maxwell_Force}
\end{equation}
where~$\boldsymbol{E}$ is the electric field and~$\rho_e$ the volumetric charge density. Neglecting magnetic effects, the electric field is irrotational,
\begin{equation}
\nabla\times\boldsymbol{E} = 0, \label{eq:Gauss_eqn}
\end{equation}
and may therefore be expressed in terms of the electrostatic potential~$\Phi$, i.e. $\boldsymbol{E} = -\nabla\Phi$. From Gauss’s law, the potential satisfies
\begin{equation}
\nabla^2\Phi = -\frac{\rho_e}{\varepsilon}, \label{eq:Poisson_Phi}
\end{equation}
where~$\varepsilon$ is the permittivity. Conservation of charge gives
\begin{equation}
\frac{\partial\rho_e}{\partial t} + \nabla\cdot\boldsymbol{J} = 0, \label{eq:phi_conservation_1}
\end{equation}
with current density
\[
\boldsymbol{J} = \sigma\boldsymbol{E} + \rho_e\boldsymbol{u},
\]
where the first term represents Ohmic conduction while the second term represents convective charge transport due to fluid motion. Thus, the conservation of charge now assumes the form
\begin{equation}
\frac{\partial\rho_e}{\partial t} + \nabla\cdot(\rho_e\boldsymbol{u}) + \nabla\cdot(\sigma\boldsymbol{E}) = 0. \label{eq:phi_conservation_2}
\end{equation}

To minimize spurious charge leakage across the interface, accurate evaluation of the interfacial values of conductivity and permittivity is crucial. Following Huh~\emph{et al.}~\cite{huh2022simulation}, a new interpolation scheme is implemented in OpenFOAM for computing cell-averaged properties in interface cells:
\begin{gather}
\sigma = \big(C_{liq}\sigma_{liq}^{1/f} + (1 - C_{liq})\sigma_{vac}^{1/f}\big)^f, \\
\varepsilon = \big(C_{liq}\varepsilon_{liq}^{1/f} + (1 - C_{liq})\varepsilon_{vac}^{1/f}\big)^f.
\label{eq:Interpolation_electrical}
\end{gather}
For $f = 1$, this expression reduces to the weighted arithmetic mean, and for $f = -1$, to the weighted harmonic mean.

\begin{figure}[ht]
\centering
\subfigure[]{\includegraphics[trim=50mm 50mm 50mm 50mm, clip, width=0.22\textwidth]{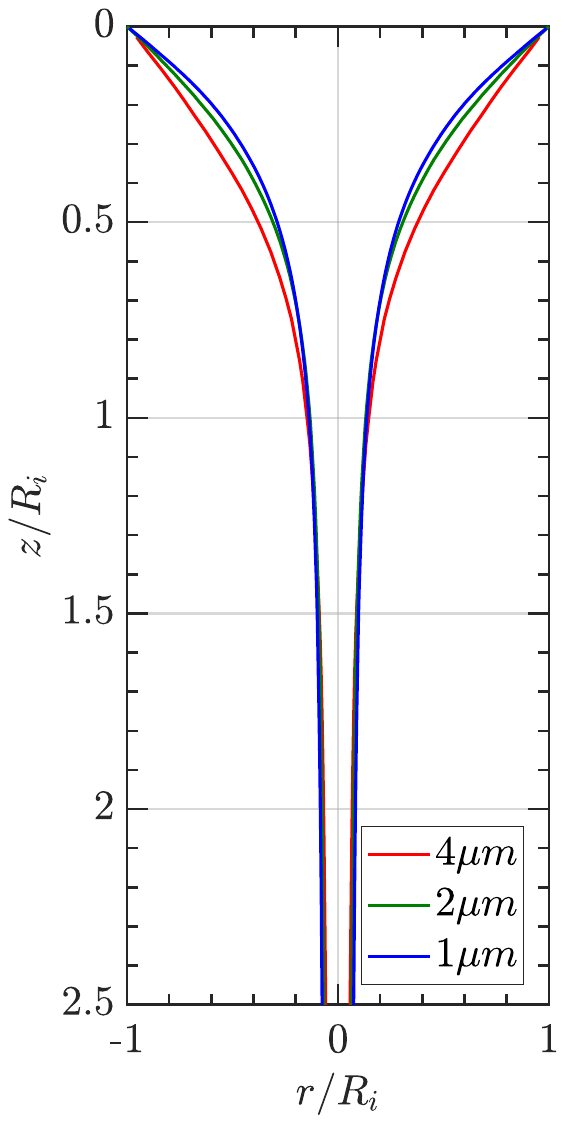}\label{fig:Gis_grid_size}}
\hspace{5mm}
\subfigure[]{\includegraphics[trim=50mm 50mm 50mm 50mm, clip, width=0.22\textwidth]{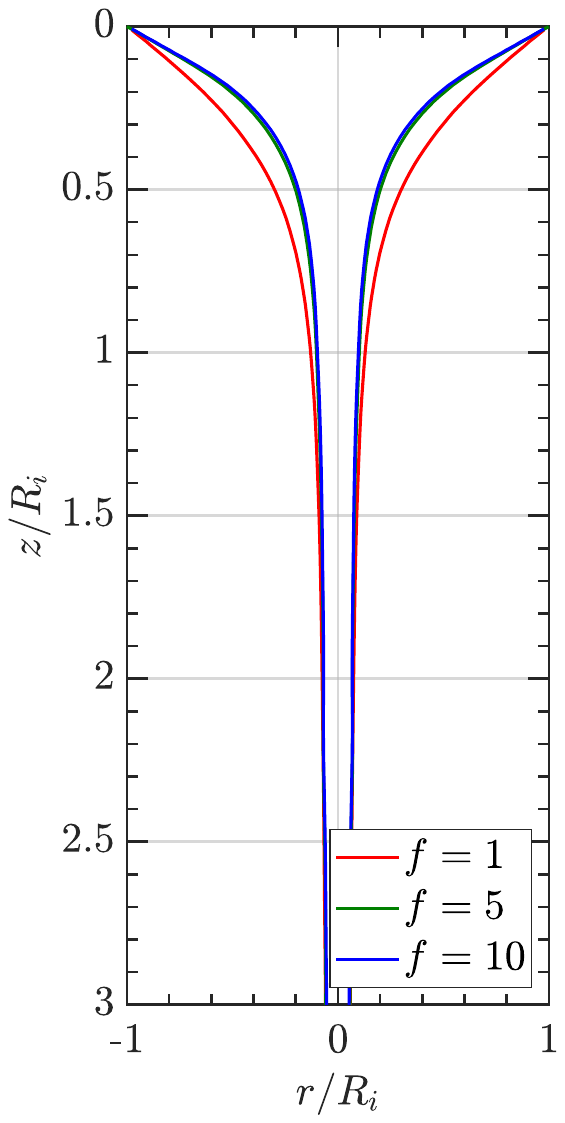}\label{fig:Gis_f}}
\caption{Grid-independence study comparing interface shapes for different mesh resolutions (a) and for varying interpolation factor $f$ (b). The study was conducted for an applied voltage of $2000~\mathrm{V}$ and a flow rate of $1~\mathrm{mL\,h^{-1}}$. The geometrical parameters are $R_o/R_i = 1.1$, $H'/R_i = 10$, and $L/H' = 1.5$. Convergence is achieved for $f = 5$ and a minimum grid spacing of $\Delta x = 2~\mu\mathrm{m}$.}
\label{fig:GIS_grid_size_f}
\end{figure}

The interpolation factor~$f$ and the finest mesh resolution were determined through a series of convergence tests, as illustrated in Fig.~\ref{fig:GIS_grid_size_f}. A structured mesh with adaptive refinement was employed, with the finest grid cells concentrated in the vicinity of the liquid–gas interface. This adaptive refinement enhances the accuracy of interface tracking and minimizes spurious charge leakage across the interface. To determine an appropriate mesh resolution, a grid-independence study was performed for an applied voltage of $2000~\mathrm{V}$ and a flow rate of $1~\mathrm{mLh^{-1}}$. Simulations were repeated with successively finer grids, and the resulting interface shapes were compared. It was found that the interface shape did not change appreciably once the smallest grid size reached approximately $2~\mu\mathrm{m}$. Accordingly, the finest grid size was fixed at $2~\mu\mathrm{m}$ in all subsequent simulations.
The interpolation factor~$f$ also influences the predicted interface shape and the local electric field distribution. To determine an optimal value of~$f$, a separate convergence study was conducted, as shown in Fig.~\ref{fig:Gis_f}. For each trial value of~$f$, the steady-state interface configuration and charge distribution were examined. Lower values of~$f$ were found to produce excessive charge diffusion across the interface, whereas excessively large values led to numerical stiffness and oscillations in the computed electric field. Convergence was achieved for $f=5$, which yielded a sharp and stable interface while maintaining good numerical stability. This value was therefore adopted in all subsequent simulations.
\begin{figure}[h]
\centering
\includegraphics[trim = 30mm 95mm 40mm 90mm, clip, angle=0, width=0.35\textwidth]{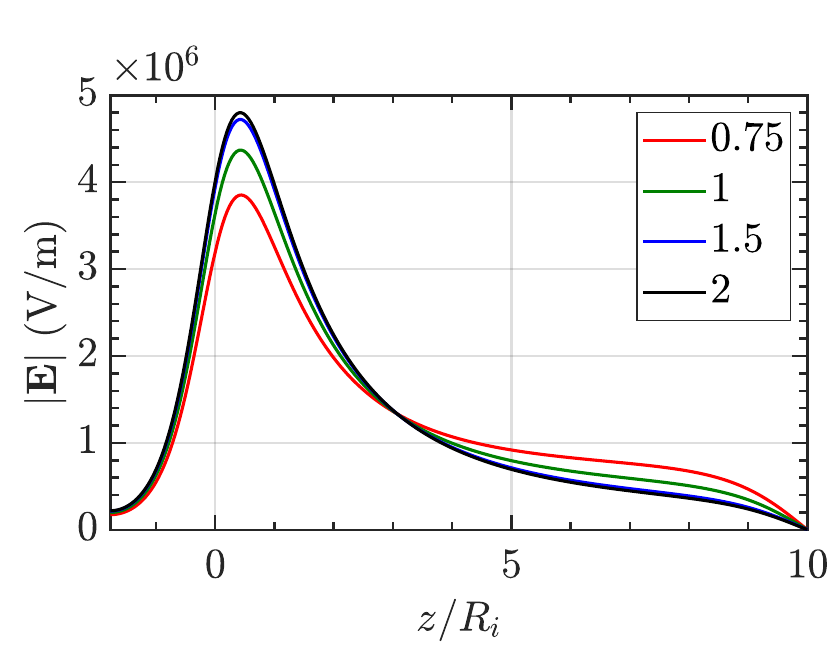}
\caption{Variation of the electric field magnitude along the needle axis for different lateral boundary positions ($L/H'$), with $H'/R_i = 10$ and $R_o/R_i = 1.1$. The needle length is $L_n/R_i = 1.77$, and convergence of the electric field is observed for $L/H' = 1.5$. All cases correspond to purely electrostatic simulations with an applied field of $E_\infty = 1.6 \times 10^6~\mathrm{V/m}$.}
\label{fig:ElectricField_convergence}
\end{figure}
Having fixed the mesh resolution and interpolation factor, it is also necessary to identify an appropriate location for the lateral boundary to minimize far-field effects. While the liquid jet remains confined to a narrow region around the axis of symmetry (see Fig.~\ref{fig:Domain_Schematic}), the electric field distribution is highly sensitive to the placement of the outer boundaries. To assess this sensitivity, a series of purely electrostatic simulations was performed by varying the radial distance~$L$ of the lateral boundary. The resulting axial electric field profiles are shown in Fig.~\ref{fig:ElectricField_convergence}. When the lateral boundary is placed closer than the tip-to-collector distance, i.e.\ $L < H'$, the peak electric field near the needle tip is underestimated. As~$L$ increases, the field distribution progressively converges, with negligible variation observed for $L \geq 1.5H'$. This analysis establishes that, in both electrostatic and electrohydrodynamic simulations, the lateral boundary should be located at least $1.5$ times the tip-to-collector distance to eliminate boundary-induced artefacts in the jet region.

\subsection{Boundary conditions without domain truncation}
To benchmark and validate the truncated-domain simulations (\emph{TDS}) proposed in this study, we first performed full-domain simulations (\emph{FDS}). The primary limitation of full-domain simulations is their computational cost, particularly when the tip-to-collector distance exceeds the needle radius by two or more orders of magnitude, i.e.\ $\mathcal{O}(H'/R_i)\sim10^2$ or larger~\cite{reneker2008electrospinning,arunachalam2024establishment,yarin2001bending}. In the present FDS, the computational domain reproduces the experimental geometry in its entirety, including a lateral boundary located at a distance of $L = 1.5H'$. The boundary conditions employed in these simulations are summarized in Table~\ref{table:boundary_conditions}, and the corresponding nomenclature is indicated in Fig.~\ref{fig:Domain_Schematic}.

The full-domain simulations are able to reproduce the cone–jet structure observed in experiments. However, capturing downstream flow features such as bending or splaying instabilities remains challenging with current CFD tools. These dynamics typically require the inclusion of additional physical effects, such as solvent evaporation, charge relaxation, or viscoelastic stresses, which are beyond the scope of the present study. Nevertheless, the simulations accurately capture the essential features of the cone–jet regime relevant to the present analysis.

\begin{table*}[t]
\centering
\small
\setlength{\tabcolsep}{15pt} 
\renewcommand{\arraystretch}{2.5}
\begin{threeparttable}
\caption{Boundary conditions for full-domain simulations.}
\label{table:boundary_conditions}
\begin{tabular}{@{} ccccccc @{}} \toprule
\thead{Boundary} & \thead{$P$} & \thead{$\mathbf{u}$} & \thead{$C$} & \thead{$\rho_e$} & \thead{$\Phi$} \\ \midrule
Inlet &
$\displaystyle\frac{\partial P}{\partial n}=0$ &
$\mathbf{u}_z = 2\,\mathbf{u}_{\text{in}}(1 - r^2)$ &
$C = 1$ &
$\displaystyle\frac{\partial\rho_e}{\partial n}=0$ &
$\displaystyle\frac{\partial\Phi}{\partial n}=0$ \\ 
Voltage wall &
$\displaystyle\frac{\partial P}{\partial n}=0$ &
$\mathbf{u}=0$ &
$\displaystyle\frac{\partial C}{\partial n}=0$ &
$\displaystyle\frac{\partial\rho_e}{\partial n}=0$ &
$\Phi = \Phi_0$ \\ 
Atmosphere &
Total pressure\tnote{*} &
$\displaystyle\frac{\partial\mathbf{u}}{\partial n}=0$ &
Inlet–Outlet\tnote{**} &
$\displaystyle\frac{\partial\rho_e}{\partial n}=0$ &
$\displaystyle\frac{\partial\Phi}{\partial n}=0$ \\ 
Ground &
$\displaystyle\frac{\partial P}{\partial n}=0$ &
$\mathbf{u}=0$ &
$\displaystyle\frac{\partial C}{\partial n}=0$ &
$\displaystyle\frac{\partial\rho_e}{\partial n}=0$ &
$\Phi = 0$ \\ 
Pressure outlet &
Total pressure &
$\displaystyle\frac{\partial\mathbf{u}}{\partial n}=0$ &
$\displaystyle\frac{\partial C}{\partial n}=0$ &
$\displaystyle\frac{\partial\rho_e}{\partial n}=0$ &
$\displaystyle\frac{\partial\Phi}{\partial n}=0$ \\ \bottomrule
\end{tabular}
\begin{tablenotes}[para,flushleft]
\footnotesize
\item[*] Total pressure is imposed through the reference pressure $p_0$ and the local velocity $\mathbf{u}$ using a fixed-value boundary condition.\\
\item[**] Inlet–outlet boundaries switch between fixed values for inflow and zero-gradient conditions for outflow.
\end{tablenotes}
\end{threeparttable}
\end{table*}

A discussion of earlier approaches for truncated-domain simulations~\cite{herrada2012numerical,ponce2018steady,lopez2023electrokinetic} and a new, simplified formulation proposed in the present study are presented in the next section.
\section{Truncated domain method}\label{sec:TDM}
Since the cone–jet occupies only a small fraction of the overall domain, the computational cost associated with simulating the cone–jet mode can be greatly reduced through the use of truncated domains. As illustrated by the red dashed lines in Fig.~\ref{fig:Domain_Schematic}, this approach confines the simulation to the local cone–jet region, thereby avoiding the need to model the entire physical domain. For such TDS, it is essential to prescribe accurate boundary conditions for all relevant field variables on the truncated outer boundaries. When appropriate boundary conditions are imposed at the truncated boundaries to account for the influence of the far-field electric field, the resulting interface dynamics can closely reproduce those obtained from full-domain simulations. One of the most widely used approaches in this regard was proposed by Herrada \textit{et al.}~\cite{herrada2012numerical}, who prescribed analytical expressions for the electric potential, $\Phi(r,z)$, at the truncated boundaries, as detailed below.

The electric potential distribution depends on the needle-to-collector distance, $H'$, and the needle radius, $R_i$. In cylindrical coordinates $(r',z')$, with the origin at the intersection of the needle axis and the grounded collector, Jones and Thong~\cite{jones1971production} developed a theoretical model for electrospraying of kerosene by approximating the needle as a semi-infinite line of charges. In the absence of free charges, the Poisson equation (equation~\eqref{eq:Poisson_Phi}) reduces to the Laplace equation, whose analytical solution is given by
\begin{equation}
\begin{alignedat}{2}
\Phi(r', z') &= {}\\
&\hspace{-2em}
-\frac{K_V \Phi_0}{\log \left(4 H' / R_{\mathrm{i}}\right)}
\log \Bigg\{
\frac{
\left[r'^2 + (H'-z')^2\right]^{1/2} + (H'-z')
}{
\left[r'^2 + (H'+z')^2\right]^{1/2} + (H'+z')
}
\Bigg\}
\end{alignedat}
\label{eq:Phi_Jones}
\end{equation}

where $\Phi_0$ denotes the applied potential on the needle, and $K_V$ is a coefficient dependent on the ratio $H'/R_i$. Using a Lagrangian formulation, Ga\~{n}\'{a}n-Calvo \textit{et al.}~\cite{ganan1994electrostatic} employed this analytical expression to study particle dynamics in sprays emitted from the cone tip. Subsequently, Herrada \textit{et al.}~\cite{herrada2012numerical}, Ponce-Torres \textit{et al.}~\cite{ponce2018steady}, and L\'{o}pez-Herrera \textit{et al.}~\cite{lopez2023electrokinetic} adopted this potential distribution to perform truncated-domain simulations of electrohydrodynamic systems.

In an axisymmetric setting, boundary conditions for $\Phi$ are prescribed on the lateral, top, and bottom boundaries of the truncated domain (see Fig.~\ref{fig:Domain_Schematic}). At the lateral boundary, $r' = R_e$, the electric potential is directly obtained from equation~\eqref{eq:Phi_Jones}, 
\begin{equation}
  \left.\Phi\right|_{\text{lateral}} = \Phi(R_e,z').
  \label{eq:Lateral_Jones}
\end{equation}
At the top boundary, $z' = H'+L_n$, where $L_n$ denotes the needle length, the potential varies only in the radial direction, and the Laplace equation simplifies to an ordinary differential equation with the solution
\begin{eqnarray}
\Phi_1(r')=\Phi_0-\left(\Phi_0-\Phi\left(R_e, H'+L_n\right)\right) \frac{\ln \left(r' / R_i\right)}{\ln \left(R_e / R_i\right)}.
 \label{eq:Top_Jones}
\end{eqnarray}
This expression is applied on the top boundary of the truncated domain for $R_o < r' < R_e$, where $R_o$ is the outer needle radius. At the bottom boundary, a Dirichlet condition cannot be directly imposed because the local electric field is strongly influenced by the presence of the liquid jet near the axis of symmetry. To address this, Herrada \textit{et al.}~\cite{herrada2012numerical} proposed a Neumann boundary condition,
\begin{equation}
  \left.\frac{d\Phi}{dz'}\right|_{z'=H'-H} = g(r'),
  \label{eq:Bottom_Jones}
\end{equation}
where $g(r')$ is obtained by differentiating equation~\eqref{eq:Phi_Jones} with respect to $z'$. Equations~\eqref{eq:Lateral_Jones}, \eqref{eq:Bottom_Jones}, and \eqref{eq:Top_Jones} together define the relevant boundary conditions for a truncated domain using the Jones and Thong approach.

\begin{figure*}
    \centering

    \subfigure[]{
        \includegraphics[trim=30mm 100mm 30mm 100mm,clip,width=0.4\textwidth]{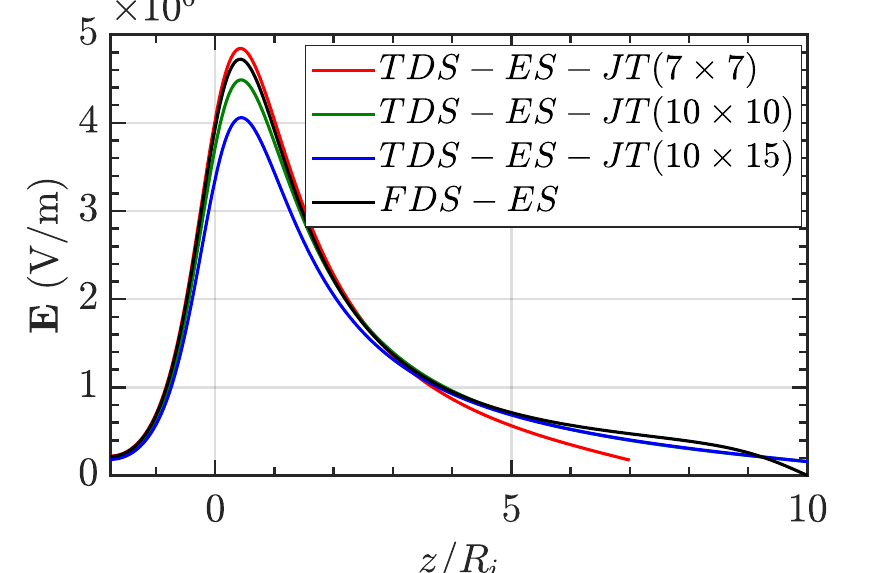}
        \label{fig:Effect_of_truncation_herrada-ES}
    }\hspace{2em}
    \subfigure[]{
        \includegraphics[trim=0mm 0mm 00mm 0mm,clip,width=0.3\textwidth]{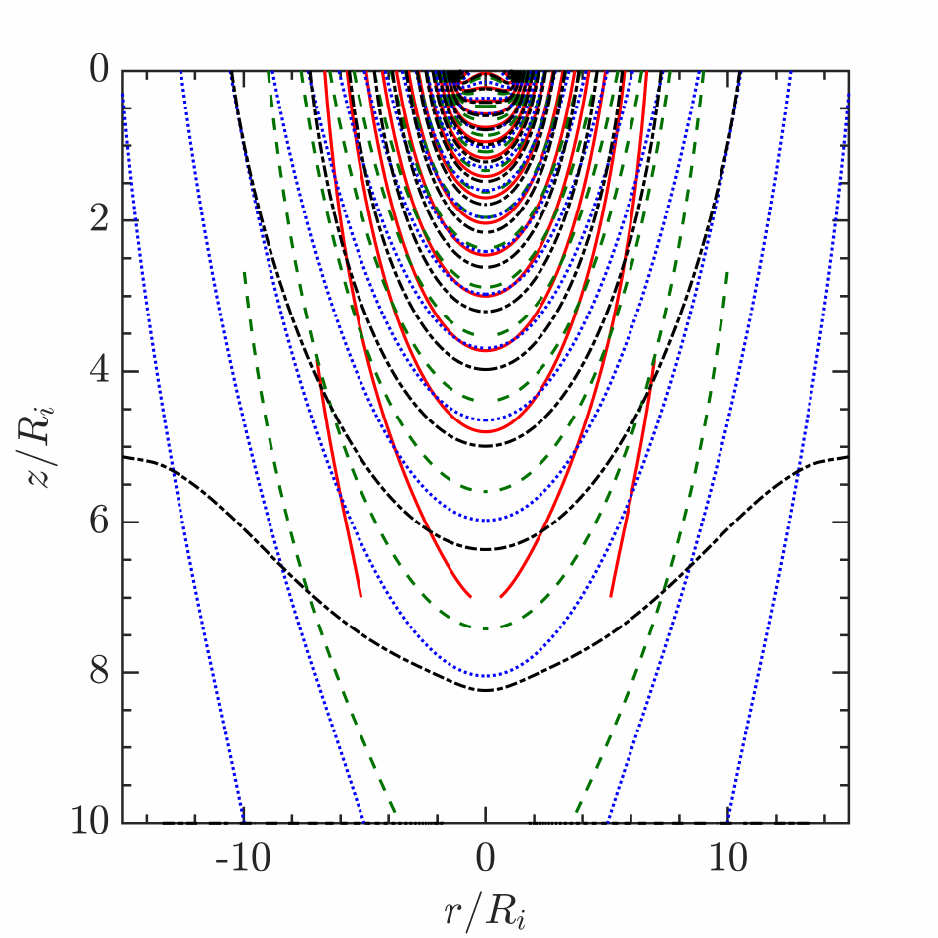}
        \label{fig:ALL_voltage_Isolines_1}
    }

    \vspace{0em}

    \subfigure[]{
        \includegraphics[trim=0mm 0mm 0mm 0mm,clip,width=0.35\textwidth]{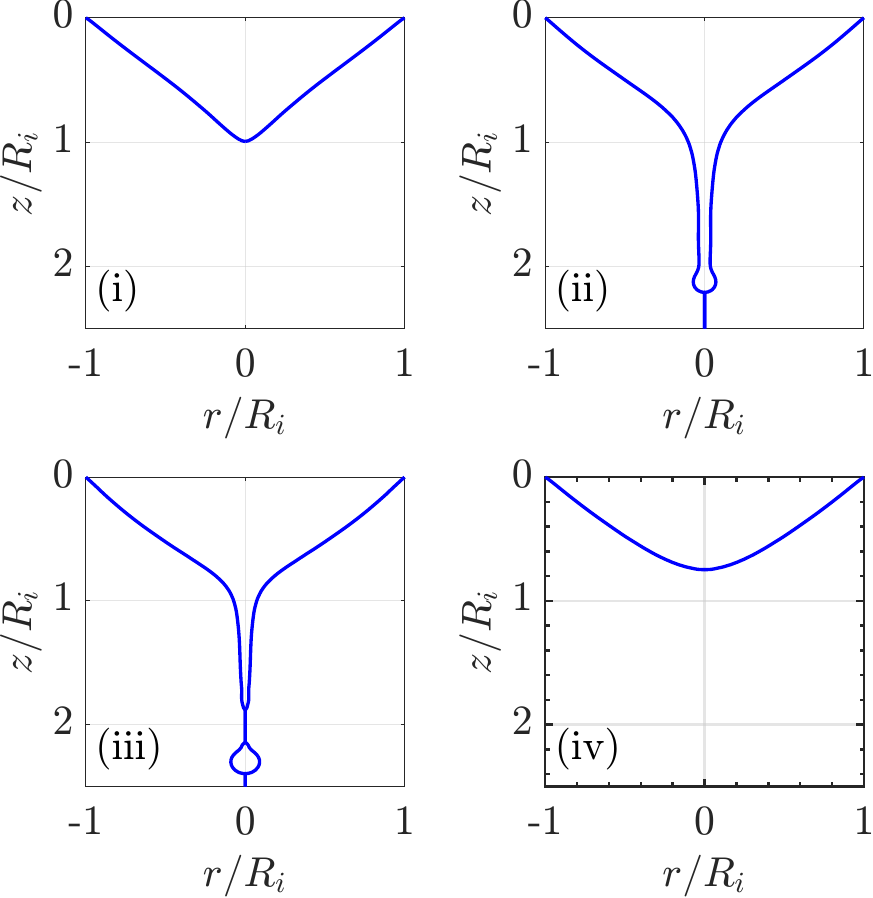}
        \label{fig:Effect_of_truncation_herrada-EHD_Dripping}
    }\hspace{5em}
    \subfigure[]{
        \includegraphics[trim=50mm 50mm 50mm 50mm,clip,width=0.2\textwidth]{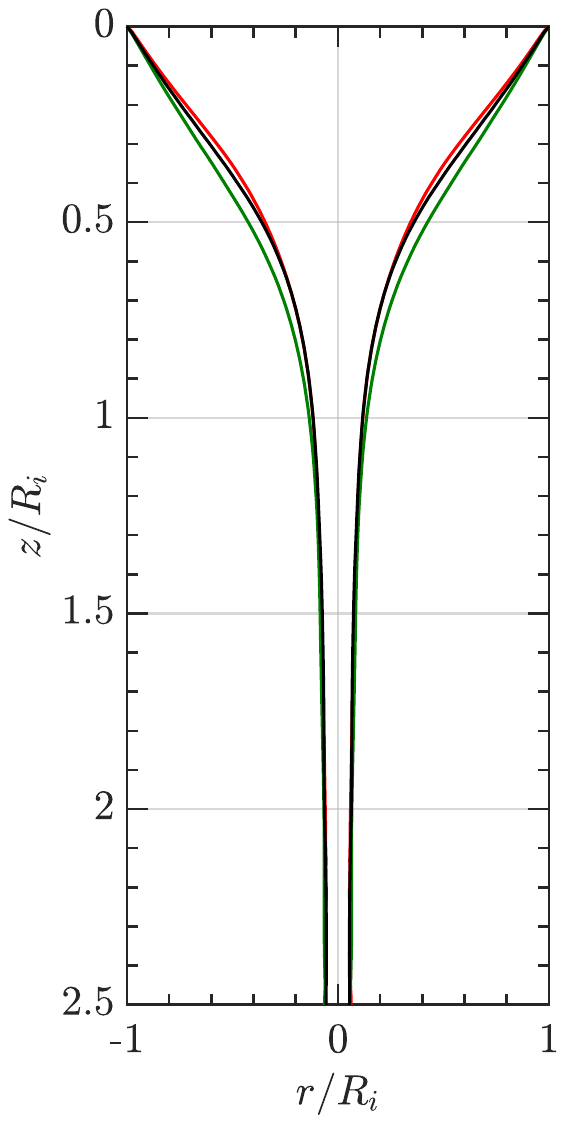}
        \label{fig:Effect_of_truncation_herrada-EHD}
    }

    \caption{(a) Variation of the electric field magnitude along the needle axis for truncated-domain electrostatic simulations (TDS-ES-JT) with three domain sizes: $7R_i \times 7R_i$ (red), $10R_i \times 10R_i$ (green), and $10R_i \times 15R_i$ (blue). The results are compared with those from the full-domain electrostatic simulation (FDS-ES, black), for which the domain size was determined from the electrostatic convergence study. 
(b) Electric potential contours for the corresponding cases, with the colour scheme consistent with that in panel (a). 
In all cases, the needle length is $L_n/R_i = 1.77$ and the outer-to-inner radius ratio is $R_o/R_i = 1.1$. The applied electric field is $E_\infty = 1.6 \times 10^6~\mathrm{V/m}$. 
(c) Interface evolution for the truncated-domain electrohydrodynamic simulation (\emph{TDS-EHD-JT}) with domain size $10R_i \times 15R_i$, showing an unstable cone–jet transitioning to the dripping mode. The sequence is shown at four time instants: $t = 3.0$, $3.2$, $3.4$, and $3.6~\mathrm{ms}$. 
(d) Interface evolution for the remaining truncated-domain cases, $7R_i \times 7R_i$ (red) and $10R_i \times 10R_i$ (green), together with the full-domain electrohydrodynamic simulation (black). The principal non-dimensional parameters used in the electrohydrodynamic simulations are listed in Table~\ref{table:Simulation_summary}. The cases shown correspond to \hyperref[case:3(a)]{Cases~3(a)--3(c)} and \hyperref[case:2(c)]{Case~2(c)}.}
    \label{fig:Under_prediction_effects}
\end{figure*}

The fidelity of any electrohydrodynamic simulation critically depends on the accuracy of the electric field distribution in the vicinity of the needle. The variation of the electric field along the axis of symmetry provides a useful measure for comparing different numerical methods. We first conduct full-domain electrostatic simulations, referred to as FDS-ES, and compare them against truncated-domain simulations employing the Jones and Thong analytical formulation, denoted as TDS-ES-JT. Fig.~\ref{fig:Effect_of_truncation_herrada-ES} presents a representative result for a specific needle length and electric field strength, $E_{\infty} = \Phi_0/H'$, for three different truncation sizes. Fig.~\ref {fig:ALL_voltage_Isolines_1} illustrates the corresponding potential contours for the full domain and the truncated domains. The variations evident in Fig.~\ref{fig:Effect_of_truncation_herrada-ES} directly influence the jet behaviour in electrohydrodynamic (EHD) simulations. Fig.~\ref{fig:Effect_of_truncation_herrada-EHD_Dripping} illustrates a dripping mode for a truncation size of $10R_i \times 15R_i$. In contrast, smaller truncation sizes of $7R_i \times 7R_i$ and $10R_i \times 10R_i$ (Fig.~\ref{fig:Effect_of_truncation_herrada-EHD}) yield stable cone–jets, albeit with slight variations in cone geometry. These truncated-domain simulations, denoted as \emph{TDS-EHD-JT}, are compared with the corresponding full-domain simulations (\emph{FDS-EHD}). It is evident that truncated-domain simulations based on the Jones and Thong approach exhibit significant sensitivity to truncation size, thereby limiting their reliability in the absence of experimental data or full-domain results for calibration.

We therefore propose a new truncated-domain scheme that employs the exact electric field distribution around the needle, obtained from full-domain electrostatic simulations. Although less elegant than the Jones and Thong analytical approach, this method guarantees convergence of the electric field for any truncation size and obviates the need for tuning the coefficient $K_V$ based on prior knowledge of the cone–jet geometry. The proposed algorithm is described below.

\subsection{Proposed method}
We introduce a robust truncated-domain methodology that overcomes the limitations of analytical potential-based approaches. The key steps are as follows:
\begin{itemize}
    \item The geometrical parameters \( H'/R_i \) and \( R_o/R_i \) are selected based on the experimental configuration, while the domain length \( L/R_i \) is determined through convergence studies to ensure accurate representation of the electric field with minimal far-field effects.
    \item A purely electrostatic full-domain simulation is performed, from which the electric field and voltage distributions are extracted along the top and lateral boundaries. Along the bottom boundary, only the normal component of the electric field gradient, $E_y = \boldsymbol{n} \cdot \boldsymbol{\nabla}\Phi$, is recorded at the chosen truncation plane.
    \item The extracted field data are smoothed and fitted with a three-term Gaussian profile to ensure continuity and differentiability:
 \begin{align}
\intertext{Top and lateral boundaries:}
\phi_{\text{fit}}
&= \sum_{i=1}^{3} a_i
\exp\!\left[-\frac{(x-b_i)^2}{2\delta_i}\right],
\label{eq:phi_fit_top} \\[6pt]
\intertext{Bottom boundary:}
\nabla \phi_{n,\text{fit}}
&= \sum_{i=1}^{3} c_i
\exp\!\left[-\frac{(x-d_i)^2}{2\xi_i}\right].
\label{eq:phi_fit_bottom}
\end{align}

    These fitted profiles are imposed as boundary conditions in the truncated domain.
    \item Finally, full electrohydrodynamic simulations are conducted within the truncated domain using these boundary conditions, providing accurate predictions with significantly reduced computational cost.
\end{itemize}

\begin{table*}[t]
    \centering
    \caption{Summary of truncated domain simulations}    
    \label{table:Simulation_summary}
    \renewcommand{\arraystretch}{1.2}
    \setlength{\tabcolsep}{6pt}
    \begin{threeparttable}

    \begin{tabular}{|c|l|c|c|c|c|c|c|c|c|c|}
        \hline
        \textbf{Case} & \textbf{Method} & $\displaystyle\mathbf{\frac{R_o}{R_i}}$ & $\displaystyle\mathbf{\frac{H'}{R_i}}$ & $\displaystyle\mathbf{\frac{H}{R_i}}$ & $\displaystyle\mathbf{\frac{L}{H'}}$ & $\displaystyle\mathbf{\frac{R_e}{R_i}}$ & $\displaystyle\mathbf{Ca_E}$ & $\displaystyle\mathbf{\tilde{\sigma}}$ & $\displaystyle\mathbf{Oh}$ & $\displaystyle\mathbf{v_e}$ \\
        \hline
        1 \label{case:1}  & Experiment \cite{herrada2012numerical} & 1.10  & 10 & --  & --   & --   & 8.54  & 17.92 & 0.172 & 0.0156 \\
        \hline
        2(a) \label{case:2(a)} & \multirow{4}{*}{Simulation: \emph{FDS-EHD}} & \multirow{4}{*}{1.10} & \multirow{4}{*}{10} & \multirow{4}{*}{--} & 0.75 & \multirow{4}{*}{--} & \multirow{4}{*}{8.54} & \multirow{4}{*}{17.92} & \multirow{4}{*}{0.172} & \multirow{4}{*}{0.0172} \\  
        \cline{1-1}\cline{6-6}
        2(b) \label{case:2(b)} &  &  &  &  & 1.00 &  &  &  &  &  \\  
        \cline{1-1}\cline{6-6}
        2(c) \label{case:2(c)} &  &  &  &  & 1.50 &  &  &  &  &  \\  
        \cline{1-1}\cline{6-6}
        2(d) \label{case:2(d)} &  &  &  &  & 2.00 &  &  &  &  &  \\ 
        \hline
        3(a) \label{case:3(a)} & \multirow{4}{*}{Simulation: \emph{TDS-EHD-JT}} & \multirow{3}{*}{1.10} & \multirow{3}{*}{10} & 10 & \multirow{3}{*}{-} & 15 & \multirow{3}{*}{8.54} & \multirow{3}{*}{17.92} & \multirow{3}{*}{0.172} & \multirow{3}{*}{0.0172} \\
        \cline{1-1}\cline{5-5}\cline{7-7}
        3(b) \label{case:3(b)} &  &  &  & 10 &  & 10 &  &  &  &  \\
        \cline{1-1}\cline{5-5}\cline{7-7}
        3(c) \label{case:3(c)} &  &  &  & 7  &  & 7  &  &  &  &  \\
        \cline{1-1}\cline{3-11}
        3(d)\textsuperscript{\textdagger} \label{case:3(d)} & & 1.22 & 40 & 9.4 & --  & 6.66 & 10.00 & 15.30 & 0.182 & 0.0182 \\
        \hline
        4 \label{case:4} & Simulation: \emph{TDS-EHD-P} & 1.10 & 10 & 6 & -- & 6 & 8.54 & 17.92 & 0.172 & 0.0172 \\
        \hline
    \end{tabular}
\begin{tablenotes}[para,flushleft]
\footnotesize
\item[\textsuperscript{\textdagger}] This case corresponds to simulations in Herrada \emph{et al.}\cite{herrada2012numerical}, but with $K_V=0.56$, which corresponds to a dimensionless tip-to-collector distance, $H'/R_i=40$, that is much larger than the experimental value.
\end{tablenotes}

\end{threeparttable}

\end{table*}

\section{Result and discussions}\label{sec:results}
Table~\ref{table:Simulation_summary} summarizes all simulations performed in the present study. The simulations are characterized by several dimensionless parameters: the Ohnesorge number, $Oh = \mu / \sqrt{\rho \gamma R_i}$; the permittivity ratio, $\varepsilon_r = \varepsilon / \varepsilon_0$; the electric capillary number, $Ca_E = \varepsilon_0 \Phi_0^2 / (R_i \gamma)$; the dimensionless conductivity, $\tilde{\sigma} = \varepsilon_r \sqrt{\rho R_i^3 \sigma^2 / (\gamma \varepsilon^2)}$; and the dimensionless velocity, $\mathbf{v}_e = Q / (\pi R_i^2 v_c)$, where $v_c = R_i / t_c$ and $t_c = \sqrt{\rho R_i^3 / \gamma}$. The simulations are grouped into three categories: full-domain electrohydrodynamic simulations (\emph{FDS-EHD}), truncated-domain electrohydrodynamic simulations based on the Jones and Thong analytical expression (\emph{TDS-EHD-JT}), and truncated-domain electrohydrodynamic simulations using the proposed method (\emph{TDS-EHD-P}).

\begin{figure}[h]
\centering
    \includegraphics[trim = 0mm 0mm 0mm 0mm, clip, angle=0,width=0.4\textwidth]{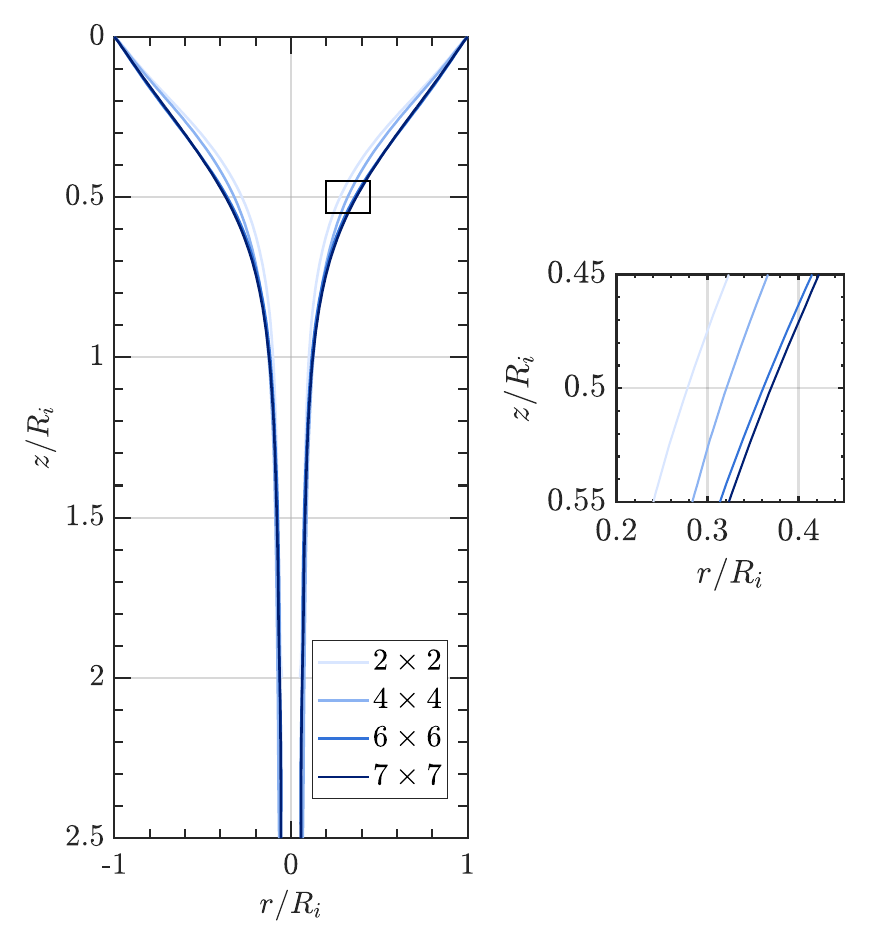}
\caption{The effect of truncation size on the interface shape is illustrated for the proposed \emph{TDS-EHD-P} method, corresponding to the non-dimensional parameters listed in \hyperref[case:4]{Case~4}. A converged interface is obtained for a domain size of $6R_i \times 6R_i$.}
\label{fig:Overall_truncated_size_convergence}
\end{figure}

To assess the efficacy of the proposed approach, we use experimental results reported by Herrada \emph{et al.}~\cite{herrada2012numerical} for electrospraying of octanol as a reference and compare the two truncated-domain approaches, \emph{TDS-EHD-JT} and \emph{TDS-EHD-P}, against full-domain simulations and experiments. The corresponding dimensionless parameters and geometrical configurations are listed in Table~\ref{table:Simulation_summary}. Cases~2(a–d) correspond to full-domain simulations performed with the same dimensionless parameters and tip-to-collector distance as the experiment (Case~1), but with four different lateral boundary locations. As shown earlier using purely electrostatic simulations (Fig.~\ref{fig:ElectricField_convergence}), a lateral boundary placement of $L/H' \geq 1.5$ yields convergence of the electric field. Consistent convergence of the cone–jet structure is also observed in full-domain electrohydrodynamic simulations for $L/H' = 1.5$ and $2.0$. We therefore adopt the results from \emph{FDS-EHD} with $L/H' = 1.5$, corresponding to Case~2(c), as the reference solution for evaluating truncated-domain simulations.

We first examine truncated-domain electrohydrodynamic simulations based on the Jones and Thong analytical expression for the electric field. This approach requires specification of two parameters: (i) the coefficient $K_V$ in Eq.~\eqref{eq:Phi_Jones}, and (ii) the size of the truncated computational domain. Jones and Thong~\cite{jones1971production} showed that $K_V$ increases with the tip-to-collector distance and that a unique value of $K_V$ may be selected for a given $H'/R_i$. For the present experiments, $H'/R_i = 10$ (\hyperref[case:1]{Case~1}), yielding $K_V = 0.182$. A convergence study was carried out for three truncation sizes: $10R_i \times 15R_i$ (\hyperref[case:3(a)]{Case~3(a)}), $10R_i \times 10R_i$ (\hyperref[case:3(b)]{Case~3(b)}), and $7R_i \times 7R_i$ (\hyperref[case:3(c)]{Case~3(c)}). Each case produced different results, reflecting the sensitivity of the electric field distribution to the truncation size, as already established from electrostatic simulations (Fig.~\ref{fig:Under_prediction_effects}). Herrada \emph{et al.}~\cite{herrada2012numerical} employed the Jones and Thong formulation with $K_V = 0.56$, corresponding to $H'/R_i = 40$, together with a truncation size of $9.4R_i \times 6.66R_i$ (\hyperref[case:3(d)]{Case~3(d)}), and obtained acceptable cone–jet shapes. However, the selection of these parameters lacks a clear physical basis and appears to rely on \emph{a priori} knowledge of the cone–jet configuration. As a result, the \emph{TDS-EHD-JT} approach cannot be readily applied to general configurations without experimental or full-domain input.

To implement the proposed truncation strategy, a convergence study was conducted by systematically varying the truncation size within the experimental parameter range. The truncation size was varied from $2R_i \times 2R_i$ to $7R_i \times 7R_i$, as shown in Fig.~\ref{fig:Overall_truncated_size_convergence}. The interface shape was found to converge for a truncation size of $6R_i \times 6R_i$, which was adopted for all subsequent simulations and corresponds to \hyperref[case:4]{Case~4}.

\begin{figure*}[t]
    \centering
    \subfigure[]{
        \includegraphics[trim=0mm 0mm 0mm 10mm, clip, width=0.24\textwidth]{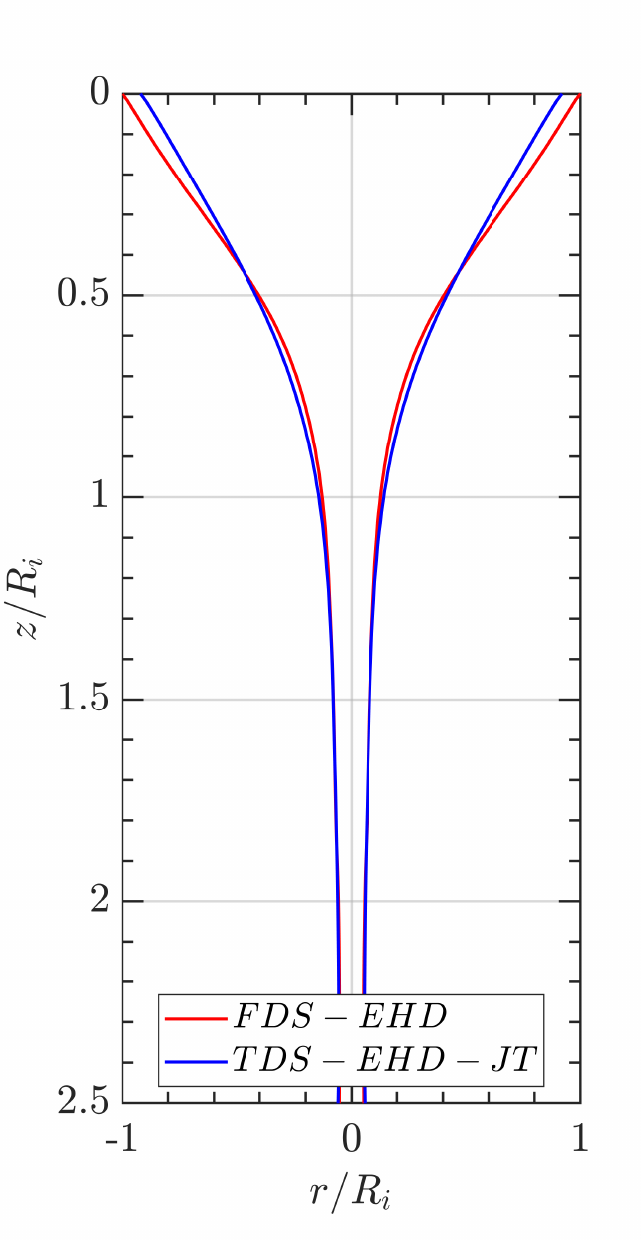}
        \label{fig:Interface_FDS_TDS_H}
    }
    \hspace{5mm}
    \subfigure[]{
        \includegraphics[trim=0mm 0mm 0mm 10mm, clip, width=0.23\textwidth]{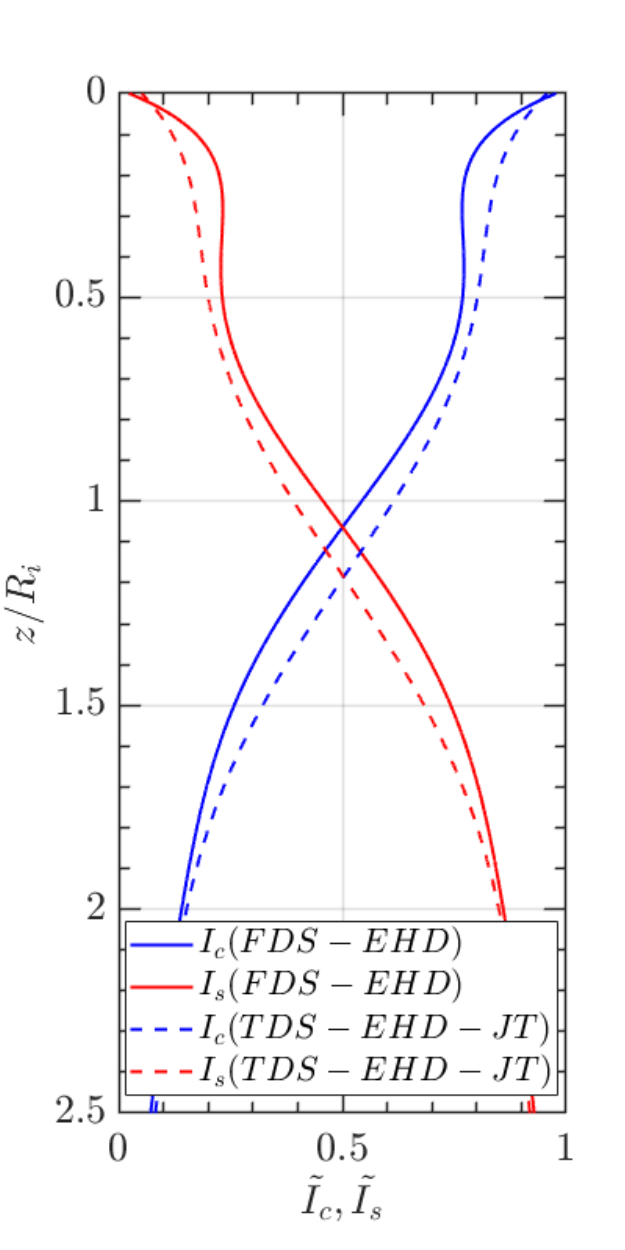}
        \label{fig:Current_FDS_TDS_H}
    }
    \hspace{5mm}
    \subfigure[]{
        \includegraphics[trim=0mm 0mm 0mm 10mm, clip, width=0.23\textwidth]{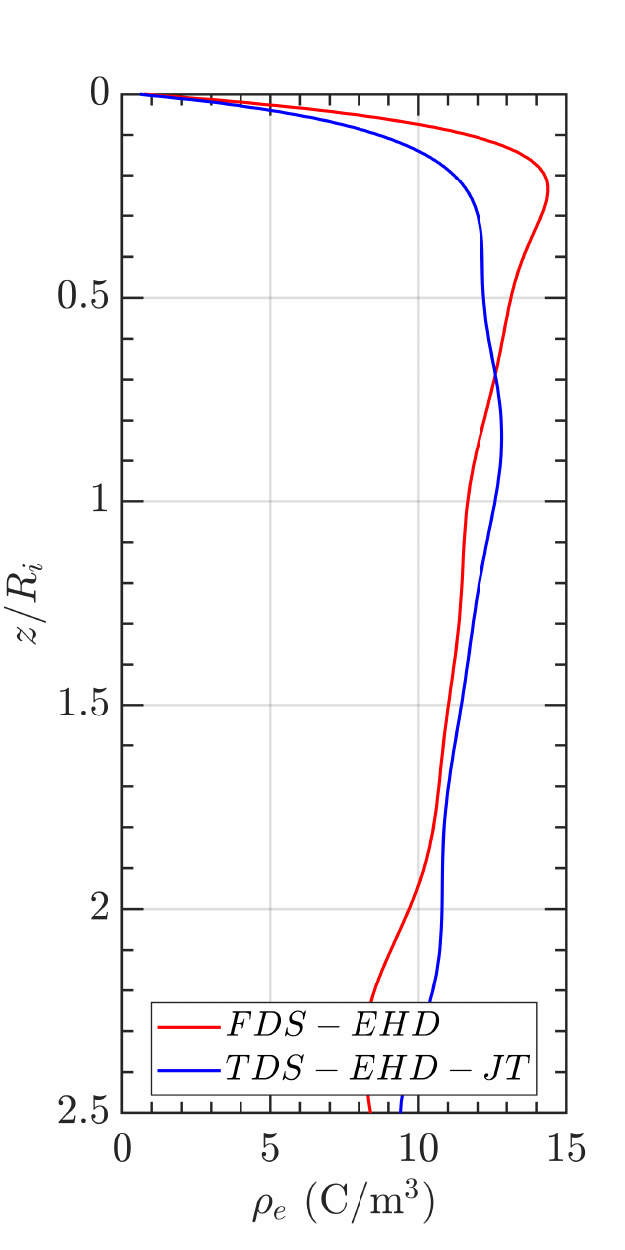}
        \label{fig:Interface_Rhoe_FDS_TDS_H}
    }
    \hspace{5mm}
    \subfigure[]{
        \includegraphics[trim=0mm 0mm 0mm 10mm, clip, width=0.23\textwidth]{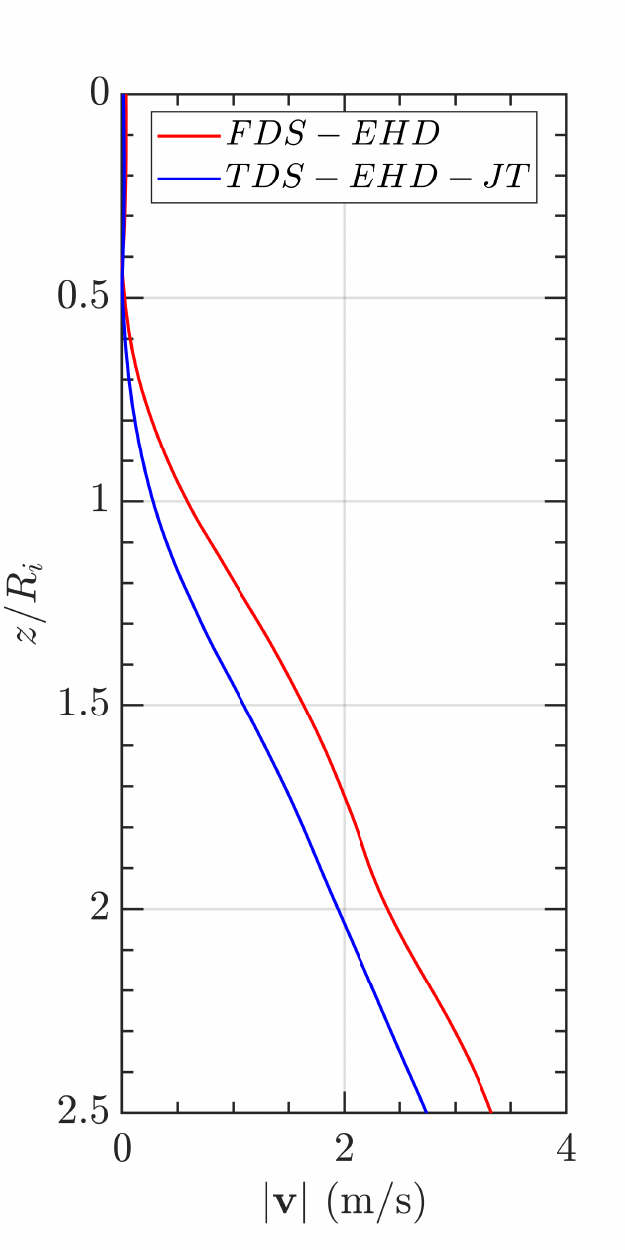}
        \label{fig:Velocity_FDS_TDS_H}
    }
     \hspace{5mm}
    \subfigure[]{
        \includegraphics[trim=0mm 0mm 0mm 10mm, clip, width=0.23\textwidth]{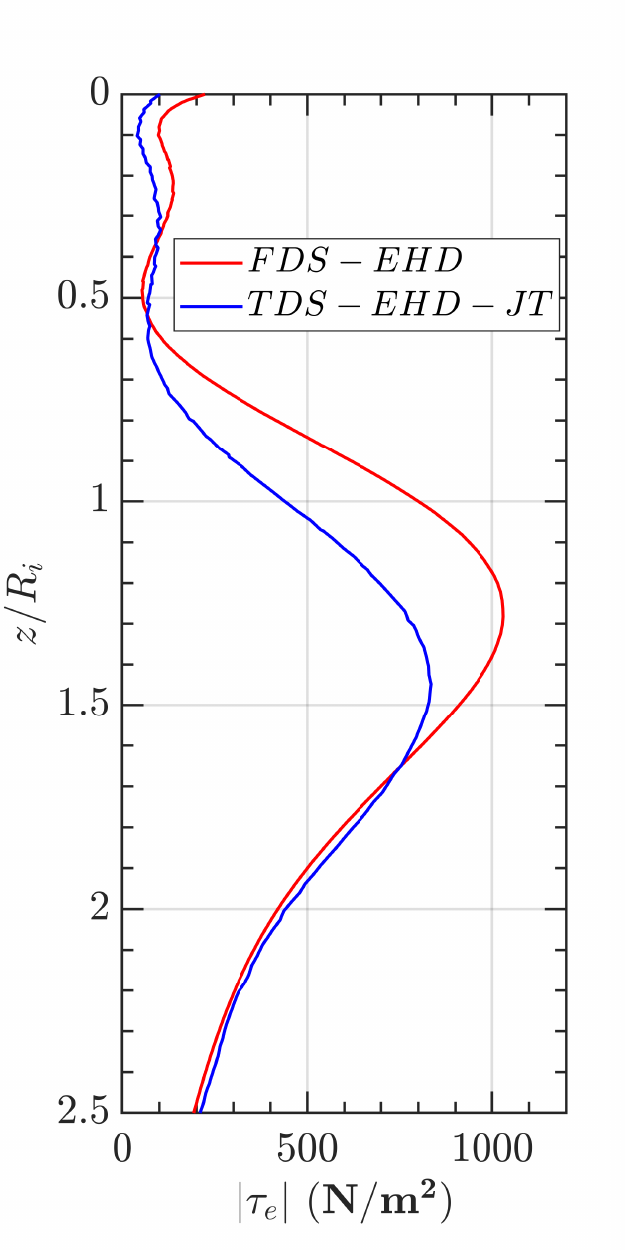}
        \label{fig:Interface_TauE_FDS_TDS_H}
    }
    \caption{(a) Comparison of the interface shapes obtained for \hyperref[case:2(c)]{Case~2(c)} and \hyperref[case:3(d)]{Case~3(d)} for 1-octanol ($\rho = 827~\mathrm{kg/m^3}$, $\mu = 0.0081~\mathrm{kg/(m\,s)}$, $\gamma = 0.0266~\mathrm{N/m}$, $\sigma = 9 \times 10^{-7}~\mathrm{S/m}$, and $\beta = 10$). The red line corresponds to \hyperref[case:2(c)]{Case~2(c)}, while the blue line corresponds to \hyperref[case:3(d)]{Case~3(d)}. 
(b) Variation of the dimensionless conduction and surface convection currents ($\tilde{I}_c$ and $\tilde{I}_s$) along the interface. 
(c) Variation of the charge density ($\rho_e$) along the interface. 
(d) Axial variation of the velocity magnitude $|\mathbf{v}|$. 
(e) Variation of the magnitude of the interfacial Maxwell stress $|\tau_e|$ along the interface. 
The needle length is $L_n = 1.77R_i$.
    }
    \label{fig:Herrada_FD_comparison}
\end{figure*}

In addition to interface morphology, full-domain simulations provide access to quantities such as conduction and convection currents, charge density, axial velocity, and Maxwell stresses. These quantities are difficult to measure experimentally but are essential for understanding the electrohydrodynamic mechanisms governing cone–jet stability~\cite{hartman1999electrohydrodynamic,dastourani2018physical}. The resulting flow structures, particularly the vortical motion and the electric field distribution, play a critical role in sustaining a stable cone–jet configuration~\cite{hayati1986mechanism}. It is therefore necessary to perform systematic comparisons between full-domain and truncated-domain simulations to assess the reliability of truncated-domain approaches. For this purpose, \hyperref[case:2(c)]{Case~2(c)}, \hyperref[case:3(d)]{Case~3(d)}, and \hyperref[case:4]{Case~4} are selected for detailed comparison.

\begin{figure*}[t]
    \centering
    \subfigure[]{
        \includegraphics[trim=50mm 50mm 60mm 60mm, clip, width=0.23\textwidth]{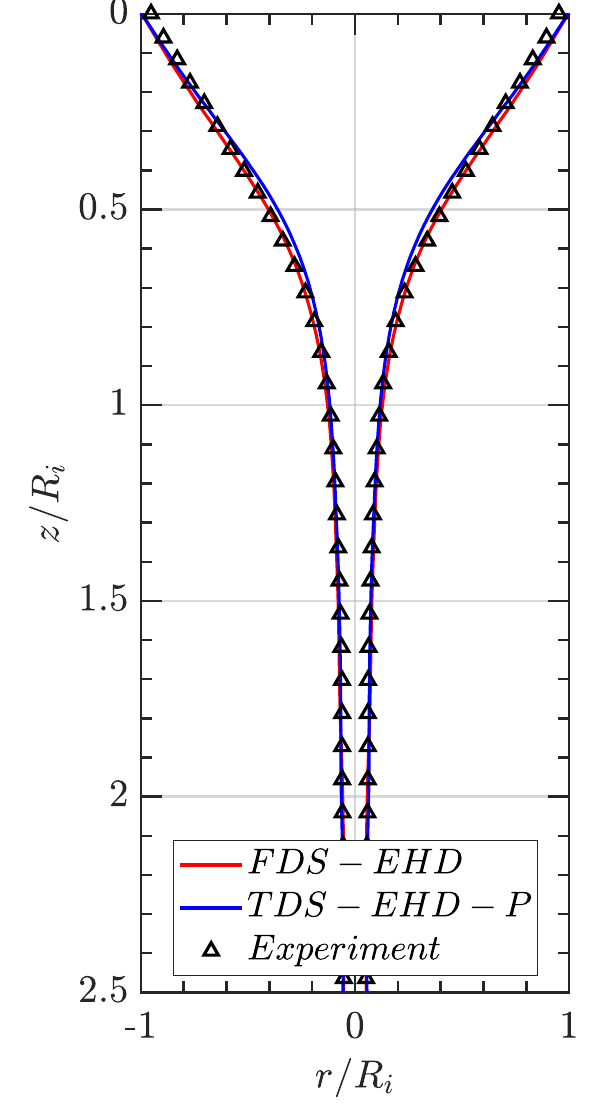}
        \label{fig:interface_FDS_TDS(P)}
    }
    \hspace{5mm}
    \subfigure[]{
        \includegraphics[trim=0mm 0mm 0mm 0mm, clip, width=0.23\textwidth]{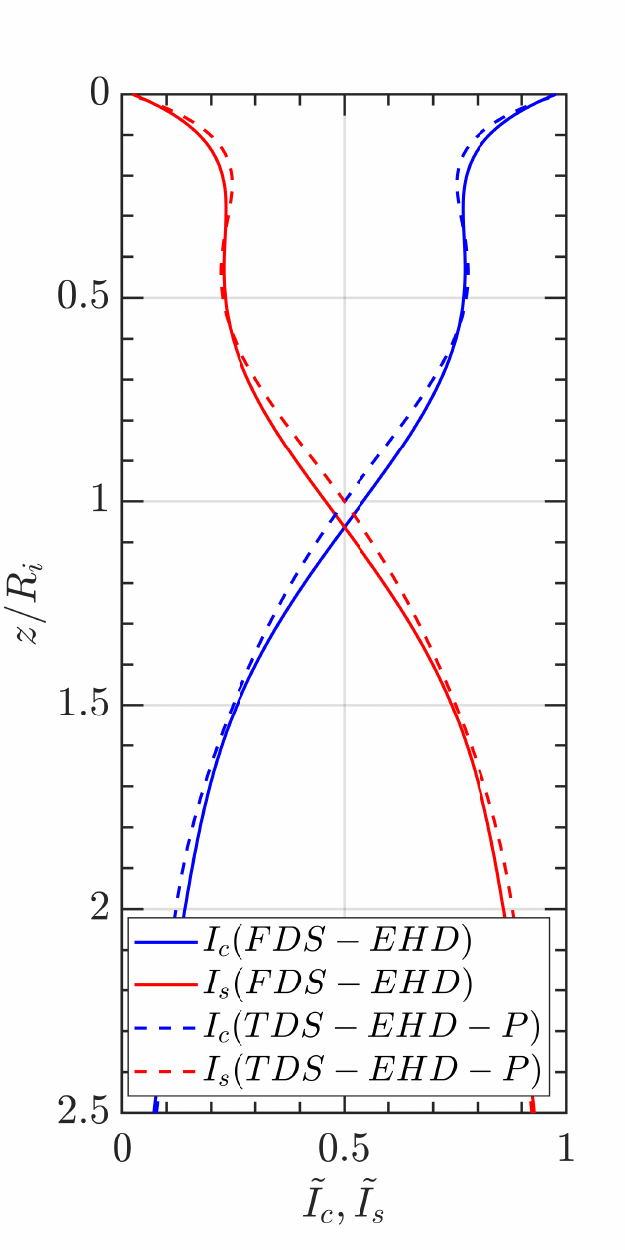}
        \label{fig:Current_FDS_TDS(P)}
    }
    \hspace{5mm}
    \subfigure[]{
        \includegraphics[trim=0mm 0mm 0mm 0mm, clip, width=0.23\textwidth]{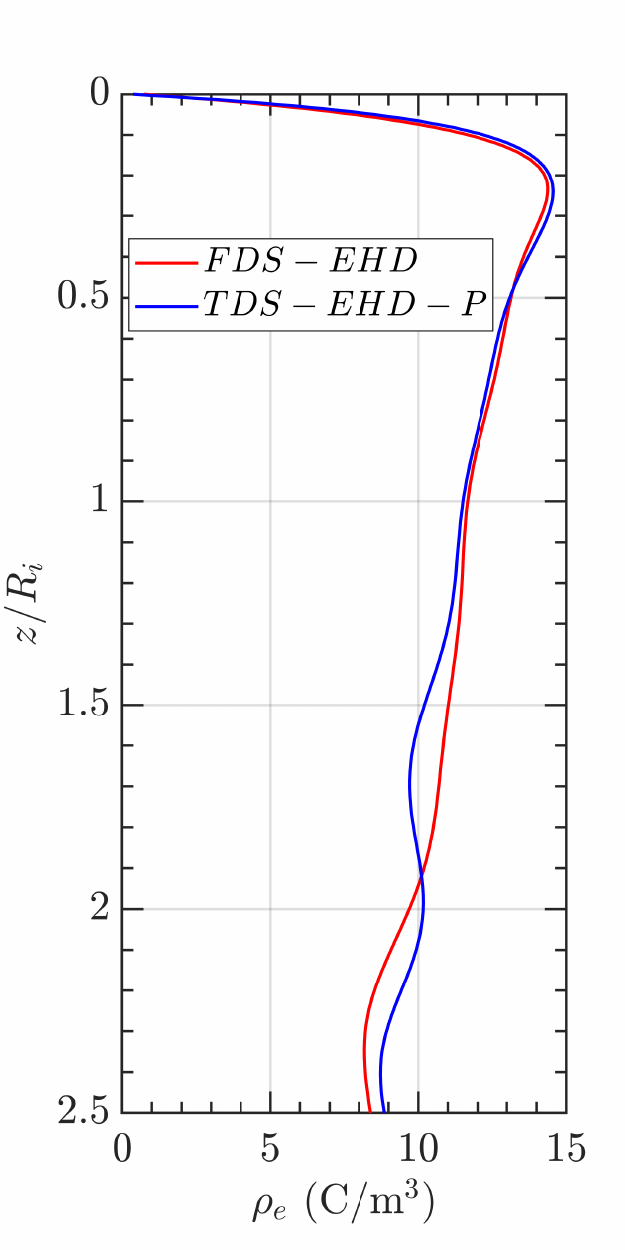}
        \label{fig:Interface_Rhoe_FDS_TDS(P)}
    }
    \hspace{5mm}
    \subfigure[]{
        \includegraphics[trim=0mm 0mm 0mm 0mm, clip, width=0.23\textwidth]{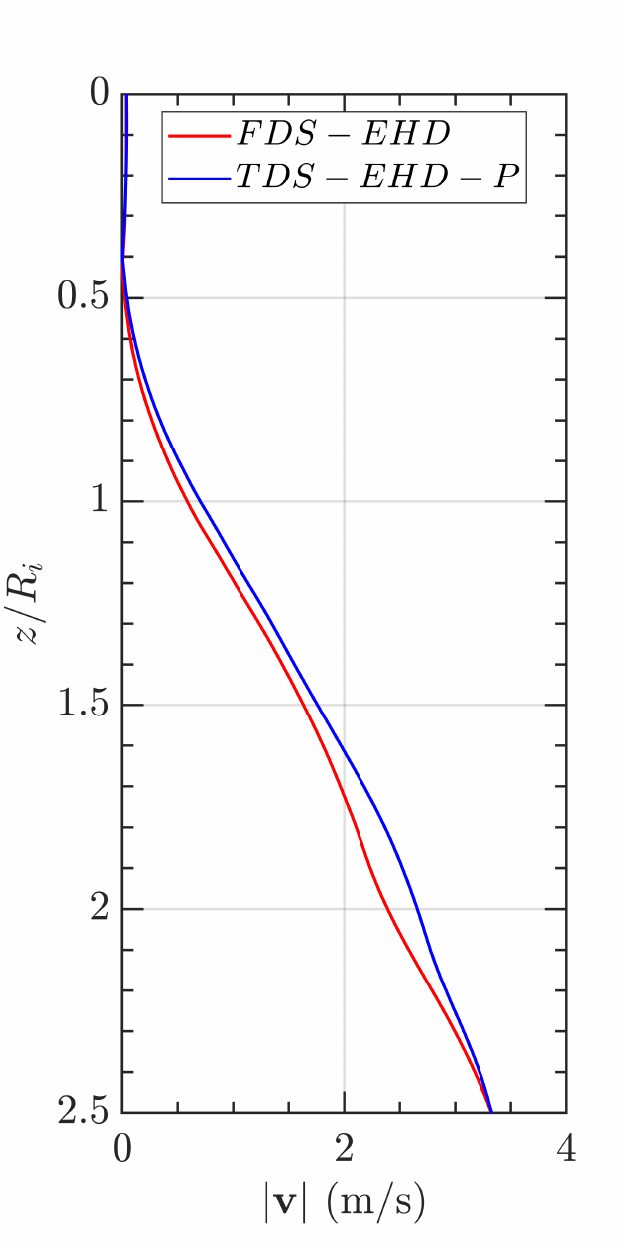}
        \label{fig:Velocity_FDS_TDS(P)}
    }
    \hspace{5mm}
    \subfigure[]{
        \includegraphics[trim=0mm 0mm 0mm 0mm, clip, width=0.23\textwidth]{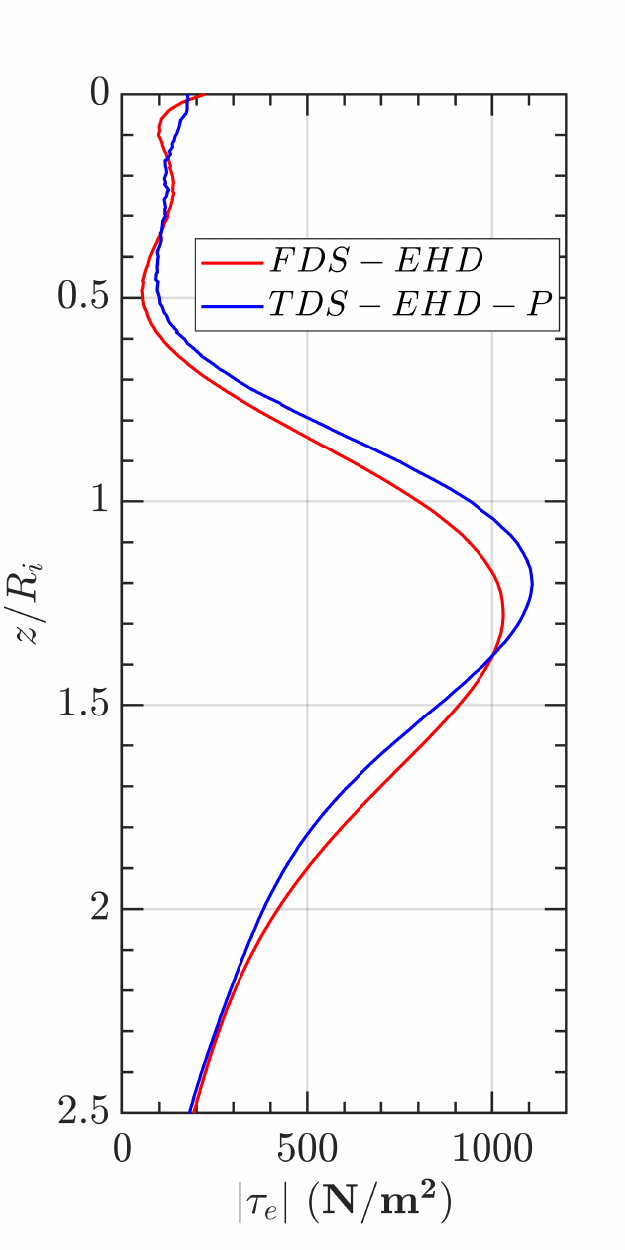}
        \label{fig:Interface_TauE_FDS_TDS(P)}
    }
    \hspace{5mm}
    \caption{(a) Comparison of the interface extracted from \hyperref[case:1]{Case~1} with simulation results from \hyperref[case:2(c)]{Case~2(c)} and \hyperref[case:4]{Case~4} for 1-octanol ($\rho = 827~\mathrm{kg/m^3}$, $\mu = 0.0081~\mathrm{kg/(m\,s)}$, $\gamma = 0.0266~\mathrm{N/m}$, $\sigma = 9 \times 10^{-7}~\mathrm{S/m}$, and $\beta = 10$). The red line corresponds to \hyperref[case:2(c)]{Case~2(c)}, the blue line corresponds to \hyperref[case:4]{Case~4}, and the symbols indicate experimental data from \hyperref[case:1]{Case~1}. 
(b) Variation of the dimensionless conduction and surface convection currents ($\tilde{I}_c$ and $\tilde{I}_s$) along the interface. 
(c) Variation of the charge density ($\rho_e$) along the interface. 
(d) Variation of the axial velocity magnitude $|\mathbf{v}|$. 
(e) Variation of the magnitude of the interfacial Maxwell stress $|\tau_e|$ along the interface. 
The needle length is $L_n = 1.77R_i$.
    }
    \label{fig:proposed_FD_comparison}
\end{figure*}

Fig.~\ref{fig:Herrada_FD_comparison} presents a comparison between full-domain electrohydrodynamic simulations (\emph{FDS-EHD}, \hyperref[case:2(c)]{Case~2(c)}) and truncated-domain simulations based on the Jones and Thong formulation (\emph{TDS-EHD-JT}, \hyperref[case:3(d)]{Case~3(d)}). As noted earlier, the physically appropriate values of $K_V$ and $H'/R_i$ do not yield a stable cone–jet regime, making direct comparison impractical. Consequently, we adopt $K_V = 0.56$ and $H'/R_i = 40$, consistent with Herrada \emph{et al.}~\cite{herrada2012numerical}. The comparison includes the cone–jet interface shape (Fig.~\ref{fig:Interface_FDS_TDS_H}), conduction and surface convection currents ($\tilde{I}_c$ and $\tilde{I}_s$, Fig.~\ref{fig:Current_FDS_TDS_H}), charge density distribution ($\rho_e$, Fig.~\ref{fig:Interface_Rhoe_FDS_TDS_H}), axial velocity field (Fig.~\ref{fig:Velocity_FDS_TDS_H}), and the magnitude of the Maxwell stress tensor (Fig.~\ref{fig:Interface_TauE_FDS_TDS_H}). Owing to the carefully chosen parameters, i.e. $K_V$ and $H'/R_i$, the interface shapes obtained from \emph{FDS-EHD} and \emph{TDS-EHD-JT} show close agreement.

Fig.~\ref{fig:proposed_FD_comparison} presents the corresponding comparison using the proposed truncation method. The agreement between the full-domain simulations and the proposed method is markedly improved for secondary quantities such as currents, charge density, and Maxwell stresses, even though the interface shape remains largely unchanged. These quantities are particularly sensitive, as they involve gradients of primary fields such as velocity and electric potential. Comparison of the current distributions in Figs.~\ref{fig:Current_FDS_TDS_H} and~\ref{fig:Current_FDS_TDS(P)} shows that the proposed method captures the spatial variation of the current more accurately. Similarly, the charge density profile obtained using the proposed method (Fig.~\ref{fig:Interface_Rhoe_FDS_TDS(P)}) closely follows the full-domain result, in contrast to the Jones and Thong–based approach (Fig.~\ref{fig:Interface_Rhoe_FDS_TDS_H}).

Importantly, the Jones and Thong approach relies on parameter tuning informed by experimental interface shapes, whereas the proposed truncation method does not require any \emph{a priori} experimental input. As a result, the proposed approach enables predictive simulations of cone–jet configurations even in the absence of experimental interface data, providing a simplified yet effective alternative to computationally expensive full-domain direct numerical simulations.

\section{Summary and Conclusions}\label{sec:conclusions}

Existing approaches for simulating the cone–jet mode in electrospinning and electrospraying rely predominantly on the analytical electric potential proposed by Jones and Thong, which has been used extensively in truncated-domain studies. In this work, we develop a systematic framework for simulating the cone–jet mode within truncated computational domains, with the objective of reducing computational cost while retaining accurate predictions of the electric field and interfacial dynamics near the needle. A series of electrostatic and electrohydrodynamic simulations was first conducted to assess the accuracy of the Jones–Thong truncation strategy. These comparisons demonstrate that the analytical potential consistently underestimates the electric field in the vicinity of the needle tip, where the field strength plays a critical role in determining the onset and stability of the cone–jet. In addition, the truncation strategy is highly sensitive to the chosen domain size: variations in the lateral or axial extent of the truncated domain lead to noticeable changes in the computed electric field, potential contours, and, ultimately, the interface profile. Although the coefficient $K_V$ in the analytical expression may be tuned to improve agreement along selected directions, such tuning does not ensure accurate reproduction of the electric field throughout the truncated region. Moreover, this tuning requires prior knowledge of the cone–jet configuration from full-domain simulations or experiments, thereby limiting the predictive capability and general applicability of the analytical approach.

To overcome these limitations, we exploit the relatively low computational cost of electrostatic simulations to obtain the correct far-field electric field distribution surrounding the needle. By performing a single full-domain electrostatic simulation, the electric potential and field distributions along the boundaries of the truncated domain are extracted and represented using compact Gaussian fits, ensuring continuity and differentiability when imposed as boundary conditions in subsequent electrohydrodynamic simulations. The resulting formulation is simple, geometry-agnostic, and eliminates the need for empirical parameter adjustment or experimental calibration.\\
An important advantage of the proposed strategy is its natural extensibility. Because no approximation is introduced in the electrostatic problem, the methodology is not restricted to axisymmetric configurations and can be easily extended to fully three-dimensional geometries. In addition, the approach is applicable to unsteady electrohydrodynamic flows, provided that the far-field electrostatic response of the system remains quasi-steady, which is typically satisfied when the timescales associated with charge relaxation and electrode charging are much shorter than the hydrodynamic timescales as shown in Appendix \ref{sec:Appendix}. Finally, although the present study demonstrates the method using a Newtonian liquid for simplicity, the formulation places no restrictions on the constitutive model for the fluid. The same strategy can therefore be readily applied to non-Newtonian or viscoelastic flows by incorporating the appropriate rheological models within the hydrodynamic solver, without modification to the electrostatic reconstruction procedure.

The performance of the proposed method was evaluated through detailed comparisons with full-domain electrohydrodynamic simulations and experimental interface profiles reported in the literature. The results show substantial improvement across all examined quantities. The cone–jet shape predicted by the proposed method closely matches both full-domain simulations and experiments. The conduction and convection current components exhibit consistent magnitudes and crossover behavior, while the charge density distributions agree closely with full-domain results throughout the cone–jet region. In addition, the velocity field and Maxwell stress profiles show markedly reduced discrepancies relative to analytical truncation-based methods. Notably, convergence of the interface shape is achieved at significantly smaller truncation sizes, leading to a substantial reduction in computational cost without loss of accuracy.

In summary, this study presents a practical and accurate approach for electrohydrodynamic simulations in truncated domains. By replacing analytical potential-based boundary conditions with those derived from inexpensive yet exact electrostatic computations, the proposed method avoids tunable parameters, does not require prior knowledge of the cone–jet shape, and achieves close quantitative agreement with both full-domain simulations and experimental observations. Its generality and simplicity make it well suited for predictive studies, parametric exploration, and optimization of electrospinning and electrospraying systems in regimes where full-domain simulations are computationally prohibitive.

\section*{Acknowledgements}\label{sec:Acknowledgements}

The author gratefully acknowledges the financial support from DST-SERB through the grants CRG/2020/003959 and CRG/2021/007096, as well as the funding received under PMRF-2002203
\vspace{-5mm}

\section*{Author declarations}\label{sec:Author declarations}
The authors have no conflicts to disclose.
\vspace{-5mm}

\section*{Data availability}\label{sec:Data availability}
The data that support the findings of this study are available within the article
\vspace{-5mm}

\appendix
\section{Parametric study - convex, concave and unsteady cone-jet shapes}\label{sec:Appendix}
\begin{figure*}[t]
    \centering

    \subfigure[]{
        \includegraphics[trim=0mm 0mm 0mm 0mm,clip,width=0.25\textwidth]{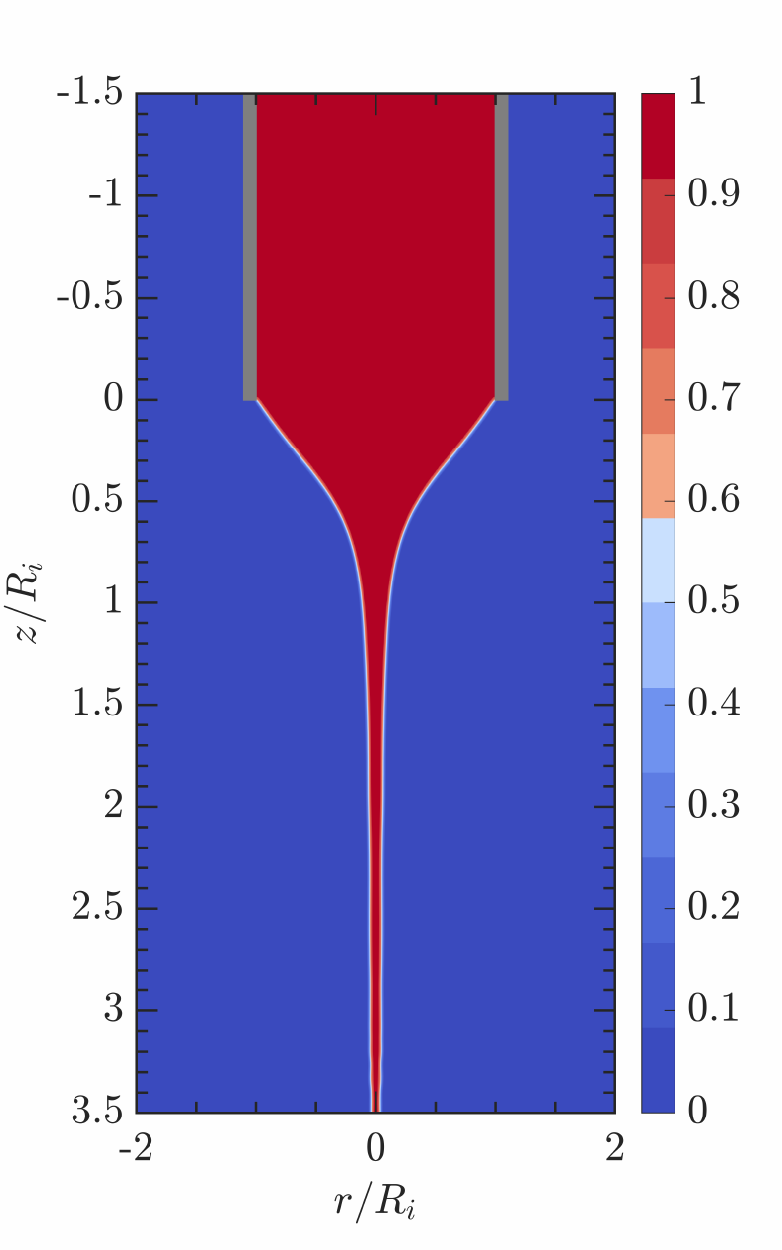}
        \label{fig:Volume_fraction_Concave}
    }
    \hspace{2em}
    \subfigure[]{
        \includegraphics[trim=0mm 0mm 0mm 0mm,clip,width=0.25\textwidth]{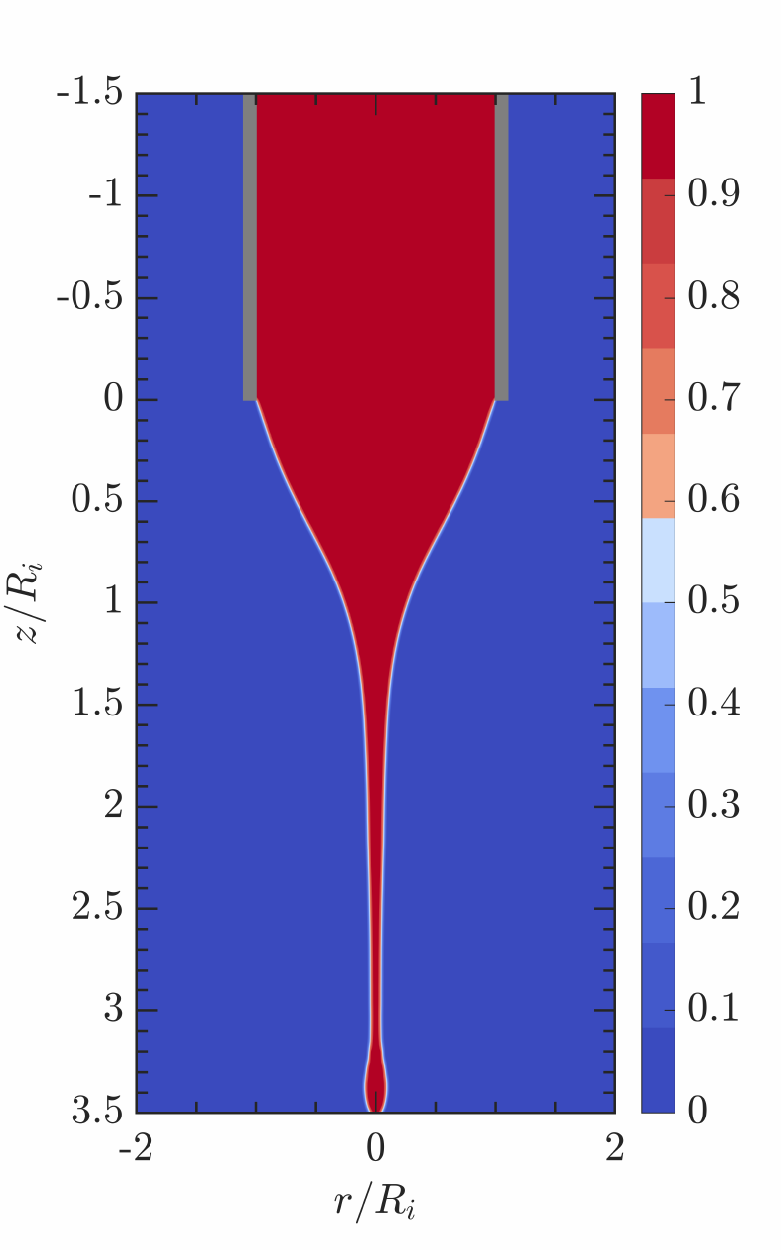}
        \label{fig:Volume_fraction_Convex}
    }
    \subfigure[]{
        \includegraphics[trim=0mm 0mm 0mm 0mm,clip,width=0.4\textwidth]{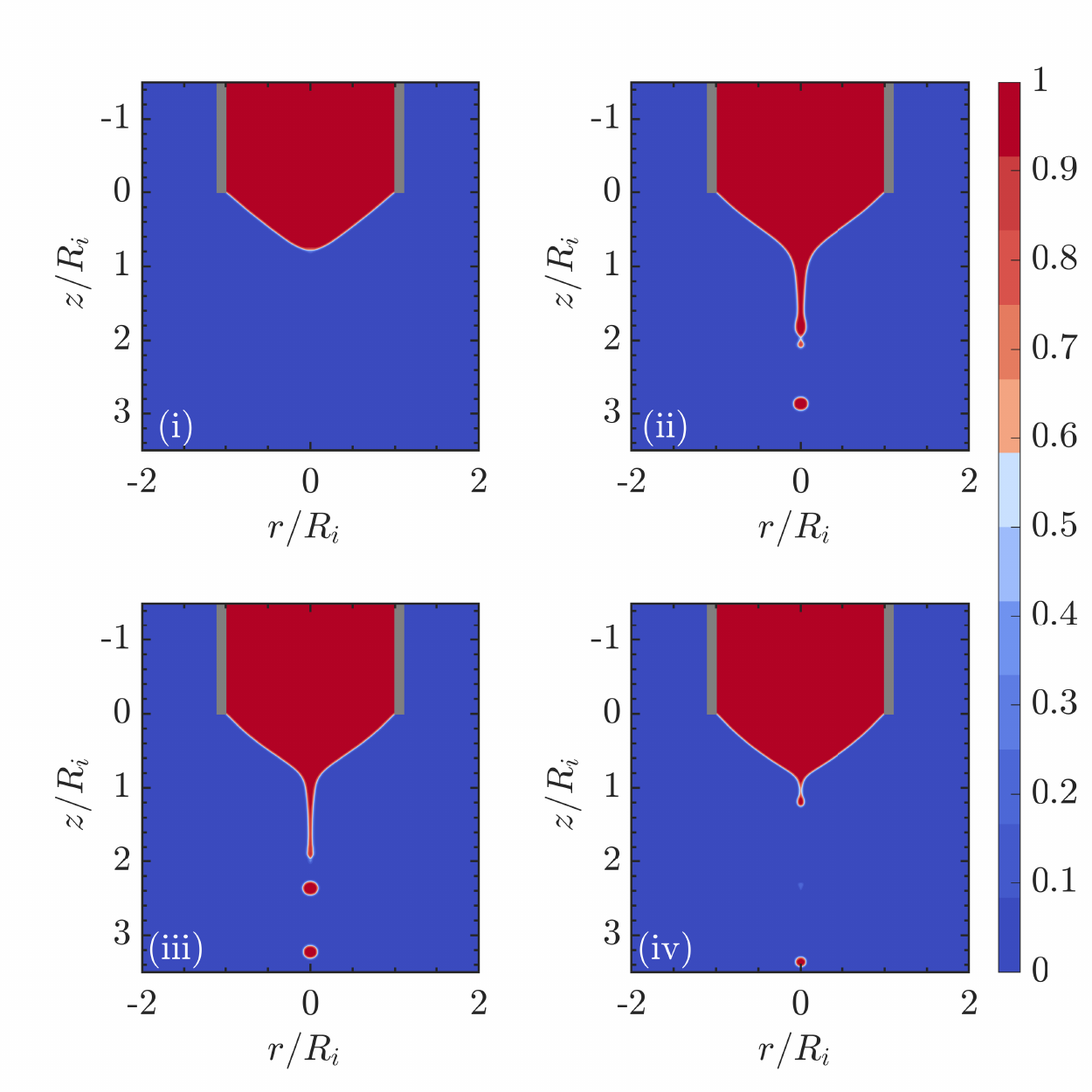}
        \label{fig:Unsteady_mode}
    }\hspace{5em}

    \caption{Spatial distribution of the volume fraction within the computational domain for different combinations of governing dimensionless parameters. The electric capillary number $\mathrm{Ca}_E$, Ohnesorge number $\mathrm{Oh}$, dimensionless conductivity $\tilde{\sigma}$, and dimensionless inlet velocity $\mathbf{v}_e$ are varied as follows: 
(a) $\mathrm{Ca}_E = 8.54$, $\mathrm{Oh} = 0.172$, $\tilde{\sigma} = 17.92$, $\mathbf{v}_e = 0.0172$; 
(b) $\mathrm{Ca}_E = 7.11$, $\mathrm{Oh} = 0.157$, $\tilde{\sigma} = 16.34$, $\mathbf{v}_e = 0.0156$; and 
(c) $\mathrm{Ca}_E = 6.26$, $\mathrm{Oh} = 0.148$, $\tilde{\sigma} = 15.40$, $\mathbf{v}_e = 0.0147$. 
Cases (a) and (b) correspond to concave and convex interface shapes, respectively, while case (c) exhibits an unsteady dripping regime. For case (c), panels (i)–(iv) show successive time instants at $t = 3.2$, $3.4$, $3.6$, and $3.8~\mathrm{ms}$, respectively. The permittivity ratio is fixed at $\varepsilon_r = 10$.
}
    \label{fig:Diffrent_Modes_and_shapes}
\end{figure*}

The solver is capable of producing steady cone–jet solutions comprising a \emph{Taylor cone} and the associated jet; however, it is not restricted to a single interface shape or operating mode. Figure~\ref{fig:Diffrent_Modes_and_shapes} illustrates the range of regimes that can be obtained using the present framework. Specifically, Fig.~\ref{fig:Volume_fraction_Concave} shows a concave interface shape, Fig.~\ref{fig:Volume_fraction_Convex} shows a convex interface shape, and Fig.~\ref{fig:Unsteady_mode} illustrates an unsteady dripping regime. 

All simulations are performed in the same computational domain, with the governing dimensionless parameters varied across the three cases. Case~(a) corresponds to $\mathrm{Ca}_E = 8.54$, $\mathrm{Oh} = 0.172$, $\tilde{\sigma} = 17.92$, and $\mathbf{v}_e = 0.0172$; case~(b) corresponds to $\mathrm{Ca}_E = 7.11$, $\mathrm{Oh} = 0.157$, $\tilde{\sigma} = 16.34$, and $\mathbf{v}_e = 0.0154$; and case~(c) corresponds to $\mathrm{Ca}_E = 6.26$, $\mathrm{Oh} = 0.148$, $\tilde{\sigma} = 15.40$, and $\mathbf{v}_e = 0.0147$. The permittivity ratio is fixed at $\varepsilon_r = 10$ for all three cases.

\section*{References}\label{sec:References}
\bibliography{Biblography}

\end{document}